\documentclass[twocolumn]{aastex631}

\usepackage{amsmath}
\usepackage{bm}
\usepackage{diagbox}\usepackage{tabularx}
\usepackage{makecell}
\usepackage{xcolor}
\usepackage{nicefrac}
\usepackage[version=4]{mhchem}

\graphicspath{{./}{Figures/}}

\defcitealias{2024arXiv240911151S}{SM25}
\defcitealias{2024ApJ...966...39S}{SM24}

\begin{document}
\title{ The Response of Planetary Atmospheres to the Impact of Icy Comets II: exo-Earth Analogues}

\author{F. Sainsbury-Martinez}
\affiliation{School of Physics and Astronomy, University of Leeds, Leeds LS2 9JT, UK}
\author{C. Walsh}
\affiliation{School of Physics and Astronomy, University of Leeds, Leeds LS2 9JT, UK}

\begin{abstract} 
The orbital regime of a terrestrial planet plays a significant role in shaping its atmospheric dynamics, climate, and hence potential habitability. The orbit is also likely to play a role in shaping the response of a planetary atmosphere to the influx of material from an icy cometary impact. 
To investigate this response, we model the impact of an icy cometary body with an Earth-analogue exoplanet (i.e. an Earth-like planet orbiting a Sun-like star with a diurnal cycle) using a cometary impact and breakup model coupled with the 3D Earth-System-Model WACCM6/CESM2. 
To quantify the role that the atmospheric dynamics play in setting the response to a cometary impact, we compare our results with a previous study investigating an impact with a tidally-locked terrestrial exoplanet.
We find that the circulation regime of the planet plays a key role in shaping the response of the atmosphere to an icy cometary impact. The weak, multi-celled circulation structure that forms on Earth-like planets is efficient at mixing material horizontally but not vertically, limiting the transport of water from the deep break-up site to higher altitudes. In turn, this limits the rate of water photodissociation at low pressures, reducing the magnitude of post-impact changes to composition. It also reduces the potential observability of an impact due to weakened cloud ice formation, and hence scattering, at low pressures. Despite this, small changes to the overall composition of the planet persist to quasi-steady-state, reinforcing the idea that ongoing bombardment may help to shape the composition/habitability of terrestrial worlds.
\end{abstract}

\keywords{Planets and Satellites: Atmospheres --- Planets and Satellites: Composition --- Planets and Satellites: Dynamics --- Methods: Computational}

\section{Introduction} \label{sec:introduction}

Material delivery by icy and rocky bodies has long been proposed as a mechanism which shapes the composition of solar-system planets. In particular, cometary and asteroidal impacts have been proposed to play a critical role in the delivery of materials, such as prebiotic \citep{oro1961,anders1989,11538074} or complex organic \citep{chyba1992} molecules, to the early Earth, helping to shape its habitability \citep{DELSEMME2000313,2020AsBio..20.1121O}. It has even been shown that microbial life can grow on meteoritic material alone \citep{Waajen2024}, reinforcing the extended role that comets/meteorites may have played in the development of life.  \\

Additionally, analysis of numerical models of the migration of small bodies from the outer solar system (where ices can form - see the review of \citealt{snodgrass2017}) inwards towards the terrestrial planets, suggests that the total mass of water delivered to the Earth is on the order of the mass of the Earths oceans \citep{2003EM&P...92...89I,2004AdSpR..33.1524I,2014Icar..239...74O,2023PhyU...66....2M}.
It has been argued that similar mechanisms should be at play in other planetary systems, with \citet{2023RSPSA.47930434A} exploring the ability of cometary impacts to deliver prebiotic molecules to rocky exoplanets, \citet{2020A&A...638A..50F} investigating the role of cometary volatile delivery in the enrichment of terrestrial planets orbiting HR 8799, and \citet{2024ApJ...966...39S} (henceforth \citetalias{2024ApJ...966...39S}) discussing the role that icy cometary impacts might have had in shaping the observed high metallicities and low C/O ratios of transiting hot Jupiters. \\

Here we build upon the work of \citet{2024arXiv240911151S} (henceforth \citetalias{2024arXiv240911151S}), who explored the effects of an icy cometary impact on the atmosphere of a tidally-locked exoplanet modelled after an Earth-like TRAPPIST-1e, extending their study to investigate the effects of a similar impact on an exo-Earth analogue, i.e. an Earth-analogue planet orbiting a Sun-like star.
In particular we are interested in quantifying how the differences in orbital regime between these two, potentially, habitable planets influences the atmospheric response to a cometary impact. Specifically, what role does the global scale nature of the circulations found in most tidally-locked atmospheres (see, for example, \citealt{2019AnRFM..51..275P,2021PNAS..11822705H}) play in the transport of impact delivered material, and how might the more localised transport in an Earth-like atmosphere which includes a day/night (diurnal) cycle differ? Of particular interest is the efficient, or otherwise, transport of water to higher altitudes, where its photodissociation plays an important role in the chemistry of the Earth's atmosphere, and where \citetalias{2024arXiv240911151S} find that cloud ice formation plays an critical role in setting the post-impact observability. 

As part of this pilot study into the effects of cometary impacts on (exo)planetary climates and atmospheric chemistry,
we couple the cometary ablation and break-up model of \citetalias{2024ApJ...966...39S} and \citetalias{2024arXiv240911151S} with WACCM6/CESM2 (\autoref{sec:method}), an Earth-System-Model which has been extensively used to explore the atmospheric dynamics of the both the Earth\footnote{see the list of nearly 3000 publications at \href{https://www.cesm.ucar.edu/publications}{https://www.cesm.ucar.edu/publications}} and exoplanets, including exo-Earths \citep[e.g.][]{10.1093/mnras/stac2604,2023MNRAS.524.1491L} and tidally-locked terrestrial planets \citep[e.g.][]{2023ApJ...959...45C,cooke2024,sainsbury2024b}. We use this coupled cometary-impact/climate model to study how the impact of a single pure-water-ice comet affects the atmosphere of an exo-Earth analogue, comparing our results (\autoref{sec:results}) with \citetalias{2024arXiv240911151S} in order to quantify the differences in atmospheric response between planets with a diurnal-cycle or fixed day/night insolation. 
We finish, in \autoref{sec:concluding}, with some concluding remarks, discussing the implications of our results for both our understanding of the composition of Earth-like exoplanetary atmospheres and the observability of icy cometary impacts, as well as possible future directions of this work.

\section{Method} \label{sec:method}

To quantify the effects of an icy cometary impact on the climate and composition of our exo-Earth analogue, we couple the parametrised cometary impact model of \citetalias{2024arXiv240911151S} (which itself is a revised version of the impact model of \citetalias{2024ApJ...966...39S}) with a modified version of the Earth-System-Model WACCM6/CESM2 { which includes} thermal energy deposition\footnote{\href{gitlab.com/leeds_work/cesm_comet}{gitlab.com/leeds\_work/cesm\_comet}},. Here we give a brief overview of the two components of this coupled model, directing readers to earlier works for a more in depth discussion. 

\subsection{Cometary Impact Model} \label{sec:cometary_impact_model}
\begin{figure}[tp] %
\begin{centering}
\includegraphics[width=0.98\columnwidth]{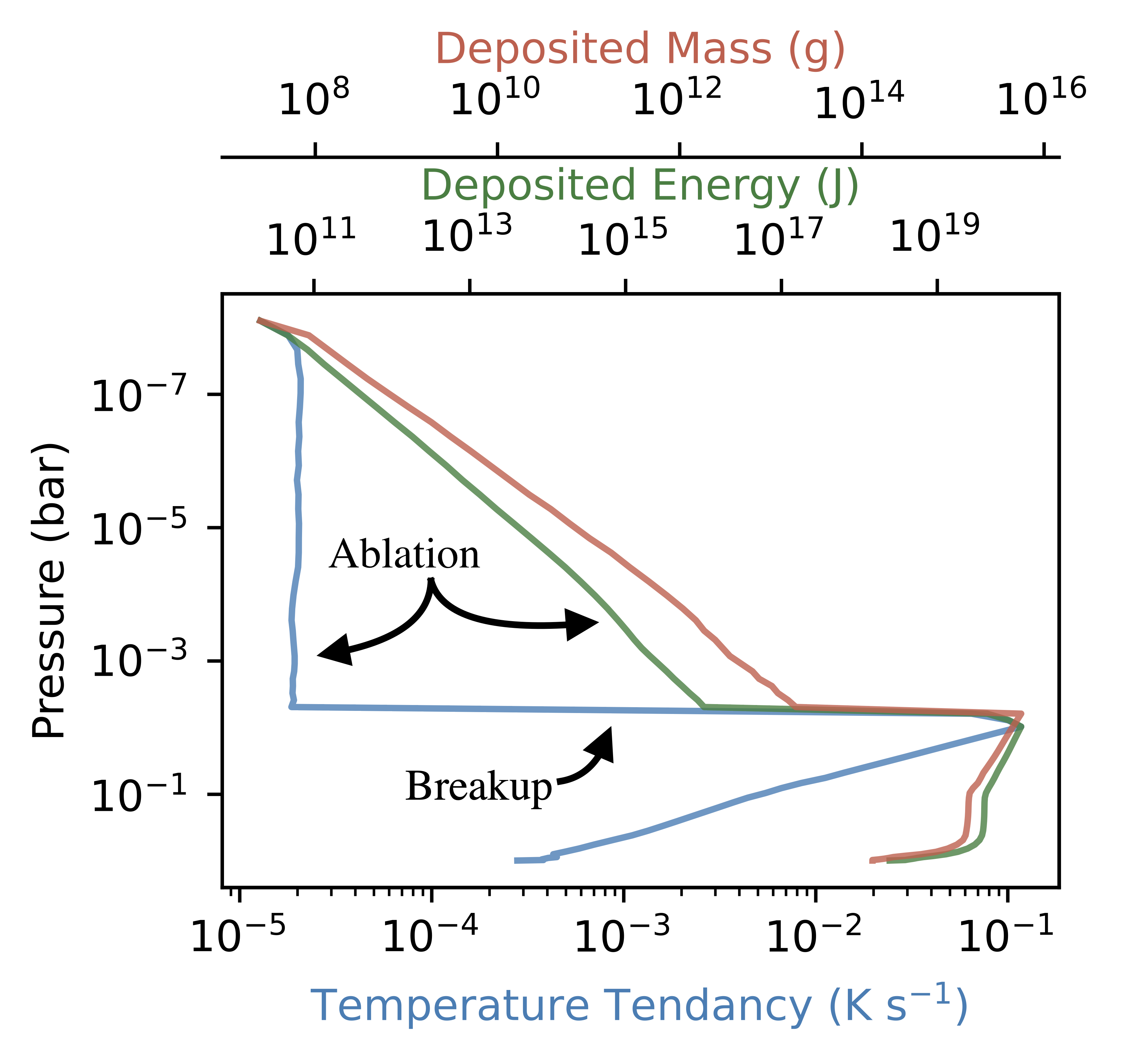}
\caption{ The initial vertical mass (red) and thermal energy distribution profile (green) generated by our cometary ablation and break-up model. Here we consider a pure-water-ice comet with a radius of 2.5 km and a density of $1\,\mathrm{g\, cm^{-3}}$, impacting over the Pacific ocean in our exo-Earth analogue atmosphere. 
\label{fig:energy_mass_deposition}  }
\end{centering}
\end{figure}
Briefly, our cometary impact model simulates the passage of a relatively low tensile strength comet through the atmosphere of a terrestrial (exo)planet. It assumes that the comet encounters the atmosphere with a zero angle of incidence ($\cos\left(\theta\right)=1$) and that it remains spherical until break-up. \\
At high altitudes, where the pressure/atmospheric-density is low, the comet slows due to atmospheric drag which drives surface {\it ablation} and hence thermally-driven mass deposition. During this ablation phase, we model the change in velocity ($dV$), mass-loss ($dM$)\footnote{Note that \autoref{eq:ablation} is incorrect in \citet{PASSEY1980211} (their equation 2), they state that the mass loss scales with $C_{H}\rho_{a}AV^{2}$ whereas it should be $C_{H}\rho_{a}AV^{3}$ as correctly stated here.}, and thermal energy deposition ($dE$) of the comet's passage in a time dependent manner (with timestep $dt$):
\begin{align}
	\frac{dV}{dt} &= g-\frac{C_{D}\rho_{a}AV^{2}}{M}, \label{eq:velocity_evo} \\
	\frac{dM}{dt} &= -\frac{C_{H}\rho_{a}AV^{3}}{2Q}\left(\frac{V^{2}-V^{2}_{cr}}{V^{2}}\right),\, \mathrm{and} \label{eq:ablation} \\
	\frac{dE}{dt} &= 0.5\pi V^{3}\rho_{a}R^{2}, \label{eq:thermal_energy}
\end{align}
where,
\begin{equation}
    A=S_{F}\left(\frac{M}{\rho_{c}}\right)^{\frac{2}{3}}
\end{equation}
is the effective cross-sectional area of the spherical comet with (remaining) mass $M$ and density $\rho_{c}$, and $S_{F}=1.3$ is the shape factor of a sphere. Additionally, in \autoref{eq:velocity_evo}, $g=9.807\,\mathrm{m\, s^{-2}}$ is the gravitational acceleration of the Earth, $C_{D}=0.5$ is the drag coefficient, $\rho_{a}$ is the local atmospheric density, and $V$ is the current velocity of the comet. In \autoref{eq:ablation}, $C_{H}=0.5$ is the heat transfer coefficient \citep{1995Icar..116..131S,2005A&A...434..343A}, $Q=2.5\times10^{10}\,\mathrm{erg\, g^{-1}}$ is the heat of ablation of the cometary ice (pure water; \citealt{2016ApJ...832...41M}), and $V_{cr}=3\,\mathrm{km\, s^{-1}}$ is the critical velocity below which no ablation occurs \citep{PASSEY1980211}. Finally, in \autoref{eq:thermal_energy}, $R$ is the current radius of the comet. \\
Note that, as discussed in \citetalias{2024arXiv240911151S}, during the ablation phase we do not directly deposit a fraction of the comet's lost kinetic energy into the atmosphere but instead we consider a reduced thermal input based upon interactions between the comet and the column of atmosphere it passes through, as described by \autoref{eq:thermal_energy}. This is done in order to maintain the stability of our atmospheric model at low pressures where the direct deposition of kinetic energy would drive massive localised heating. \\

As the comet continues its passage through the atmosphere, and the local atmospheric density increases, the drag and stresses acting on the comet also rise, leading to an increase in the ram pressure ($P_\mathrm{ram}$) and the eventual breakup of the comet. Here, we consider breakup to have occurred when the ram pressure exceeds the tensile strength of the comet ($\sigma_{T}=4.6\times10^{6}\,\mathrm{erg\, cm^{-2}}$, which is that of an icy planetesimal taken from \citealt{2016ApJ...832...41M}),
\begin{align}
    P_\mathrm{ram} &> \sigma_{T},
\end{align}
where,
\begin{align}
    P_\mathrm{ram} &= C_{D}\rho_{a}V^{2}. 
\end{align}
Once this condition is fulfilled, the passage of the comet through the atmosphere is considered to be complete. Any remaining mass and kinetic energy is then distributed deeper into the atmosphere using a scaled exponentially decaying function in order to mimic the rapid breakup of the comet, and the resulting mass/energy distribution due to the inertia of the impacting material. \\

We consider the impact of a $R=2.5\,$km pure-water-ice comet (with $\rho=1\,\mathrm{g\, cm^{-3}}$) comet, travelling with an initial velocity of $V_{0}=10\,\mathrm{km\, s^{-1}}$ (which is approximately the escape velocity of the Earth), with an unimpacted reference atmosphere based upon a pre-industrial model of the Earth (see \autoref{sec:atmosphere_model}). \\
The calculated mass and thermal energy deposition profiles are shown in \autoref{fig:energy_mass_deposition}. Here the two stages of our cometary impact, ablation and breakup, and the difference in mass/energy deposition between them, can be seen. In the low pressure/density outer atmosphere we find that both the ablation and thermal energy deposition rates increase with pressure/density. Yet they remain small relative to the mass/thermal-energy deposition rates post breakup, which occurs at around $5.46\times10^{-3}$ bar, or about $33$ km above the surface. As such, our impact model suggests that the majority of the cometary material, and kinetic energy, is delivered relatively deep into the atmosphere. 

\subsection{Exo-Earth Analogue Atmospheric Model} \label{sec:atmosphere_model}

To quantify the effects of the inclusion of a diurnal cycle on the response of a terrestrial (exo-Earth analogue) atmosphere to an icy cometary impact, we couple the above cometary ablation and breakup model with an Earth-analogue atmospheric model calculated using WACCM6/CESM2. We then run this fiducial, impacted, model, to quasi-steady-state, along with a companion, non-impacted, reference atmosphere. 

\subsubsection{WACCM6/CESM2}
The Whole Atmosphere Community Climate Model (WACCM6) is a well documented \citep{https://doi.org/10.1029/2019JD030943}, high-top (the atmosphere extends to 140 km - $\sim10^{-8}$ bar - above the surface), configuration of the open-source Coupled Earth-System Model (CESM2). It includes a modern, Earth-like, land-ocean distribution (including orography), and numerous initial atmospheric compositions, such as the pre-industrialisation, Earth-like, composition we consider here. Horizontally, the model has a resolution of $1.875^{\circ}$ by $2.5^{\circ}$, corresponding to 96 cells latitudinally (north-south) and 144 cells zonally (east-west). Vertically the model is split into 70 pressure levels distributed in $\log\left(P\right)$ space such that the number of pressure levels increases near the surface. It has been modified by \citetalias{2024arXiv240911151S}\footnote{see also \url{gitlab.com/leeds work/cesm comet}} to include two additional external forcing terms which correspond to the cometary material delivery: water and thermal energy. Both of these external forcings take the form of a rate of material or thermal energy input as well as a time-frame over which to apply said input. In both cases, to maintain model stability, we spread the deposited material/energy out both spatially, over nine-columns surrounding the impact location\footnote{over the pacific ocean at the equator, i.e., at a longitude of 180$^\circ$ and latitude of 0$^\circ$}, and temporally, over ten days of simulation time. Additionally, we convert the deposited thermal energy from a energy flux to a temperature tendency { (i.e., a rate of temperature change), which takes the form,
\begin{align}
	\frac{dT}{dt} = \frac{dE}{dt} \cdot \frac{1}{\rho_{a}\mathcal{V}c_{p}},
\end{align}
where $\mathcal{V}$ is the volume of the cell in which the thermal energy has been deposited, and $c_{p}$ is the specific heat capacity at constant pressure}. And which we show in blue in \autoref{fig:energy_mass_deposition}. 

\subsection{exo-Earth}

As our initial and unperturbed reference state, we consider the Pre-Industrial reference Earth atmospheric model which is distributed as part of the CESM 2.1 model release. We use the BWma1850 component set\footnote{A full list of component sets can be found at \url{https://www.cesm.ucar.edu/models/cesm2/config/2.1.3/compsets.html}} which couples atmospheric chemistry (which includes 98 species, 208 chemical reactions, and 90 photolysis reactions), with a dynamic ocean and a modern, i.e. Earth-like, land-ocean configuration including orography. Both the short-wave and UV irradiation of the atmosphere are based on the modern-day solar irradiation at the Earth's current orbital separation. \\
{ Similar to} the tidally-locked atmospheric models of \citetalias{2024arXiv240911151S}, we find that our icy cometary impact perturbs the multi-year atmospheric oscillations leading to variations that are out of sync with our unperturbed reference state (see the annual mean temperature and water abundance profiles shown in \autoref{fig:combined_time_evolution} in \autoref{sec:appendix}). Hence we run our impacted atmospheric model to a quasi-steady-state in which the inter-annual variation of both the mean temperature and water abundance is { qualitatively} similar to that found in our reference state. { Quantitatively, we determine that our impacted atmosphere has reached a quasi-steady-state when the difference in, mid-atmosphere, fractional water abundance between our impacted and reference atmospheres reaches a minima/plateaus (see \autoref{fig:MA_water_diff} in \autoref{sec:appendix}).} For the impact modelled here, our model has reached a quasi-steady-state $\sim20$ years post-impact.

\section{Results} \label{sec:results}
Here we explore how the water and thermal energy deposition associated with a $2.5$ km radius pure-water-ice comet impacting the atmosphere of an Earth-like planet orbiting a Sun-like star (our so-called exo-Earth analogue) affects the atmospheric composition and climate (\autoref{sub:water_abundance_and_mean_temperature} and \autoref{sub:effects_of_water_on_composition}). We also explore how both local and global circulations shape this response (\autoref{sub:advection_of_water}), comparing our results with the tidally-locked impact model of \citetalias{2024arXiv240911151S}. Finally we address the potential observability of our isolated impact using synthetic transmission, emission, and reflected light spectra in \autoref{sub:observational_implications}. \\
Furthermore, in \autoref{sec:appendix}, we perform some additional comparisons with the tidally-locked, TRAPPIST-1e-like, impact model of \citetalias{2024arXiv240911151S}, focusing on the time evolution of the atmospheres water enrichment and the formation of clouds and cloud ice. 

\begin{figure*}[tbp]
\begin{centering}
\includegraphics[width=0.84\textwidth]{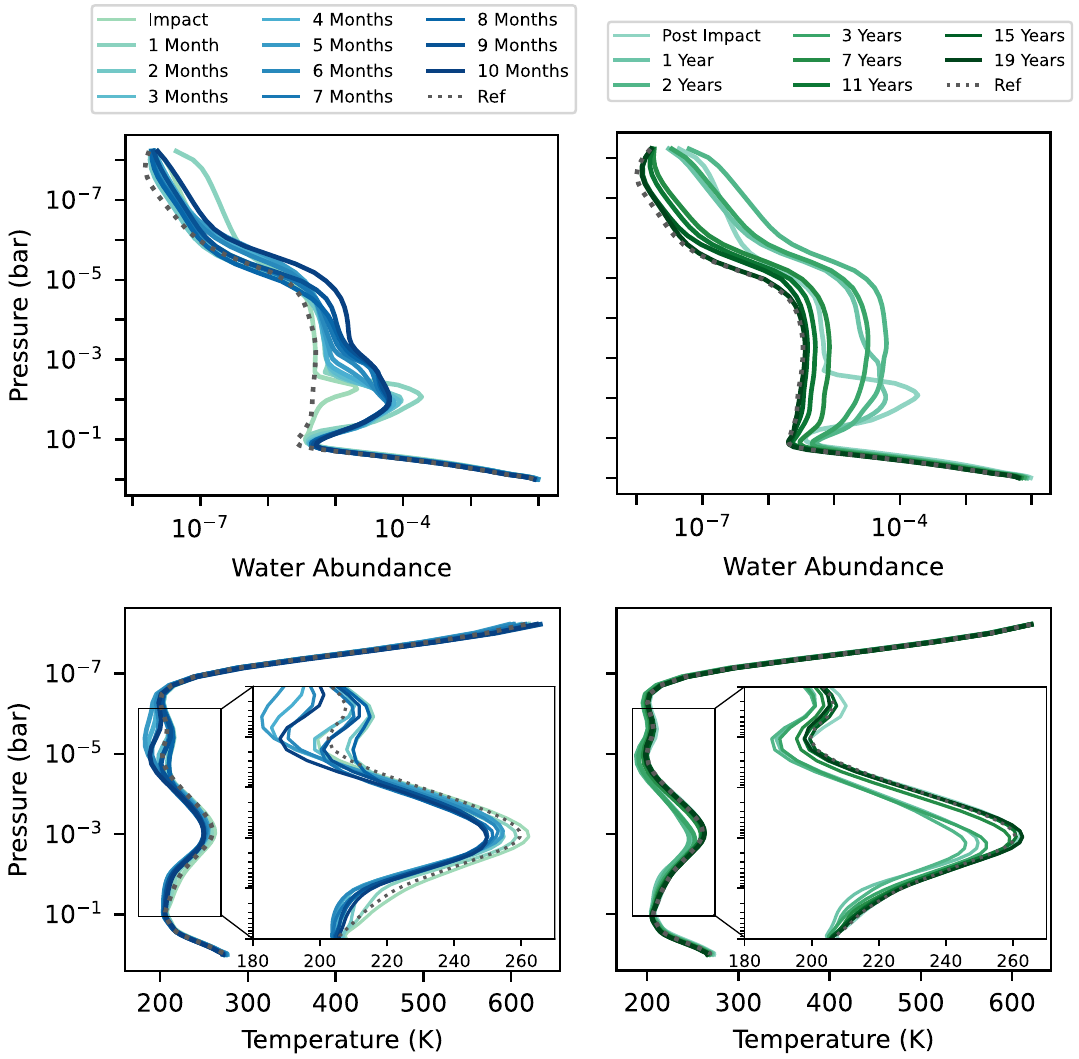}
\caption{ Fractional water (\ce{H2O}) abundance (top) and temperature (bottom) profiles calculated during the first ten months post-impact (left) and over the approximately twenty years required for the atmosphere to settle into a quasi-steady-state (right) reminiscent of our reference state (grey). Each profile is calculated by averaging both horizontally over all latitudes and longitudes, and temporally over a month of simulation time. To better demonstrate the cooling of the mid-atmosphere, we include an inset showing the temperature profile between $10^{-2}$ and $10^{-5}$ bar.    \label{fig:combined_curves_water_temp} }
\end{centering}
\end{figure*}
\subsection{Water Abundance and Mean Temperature} \label{sub:water_abundance_and_mean_temperature}
We start by exploring the direct effects of the cometary impact on the atmospheric water abundance and thermal properties (i.e., temperature). As shown in \autoref{fig:combined_curves_water_temp}, we investigate the time variations of the horizontal mean (i.e, mean over all longitudes and latitudes) fractional water abundance (top) and atmospheric temperature (bottom) both shortly after impact (left) and over the $\sim20$ years required for the atmosphere to reach a quasi-steady state (right). Note that the increase in abundance between the impact profile and the 1 month post-impact profile (henceforth referred to as `Post-Impact') occurs because we are plotting the monthly mean values and the impact occurs part-way though the month of February, leading to a reduced, temporally averaged, value.  \\

The top row of \autoref{fig:combined_curves_water_temp} reveals both how a solitary impact can affect the overall water content of the atmosphere as well as the key role that vertical mixing plays in setting the fractional water abundance profile. Note that the term fractional abundance refers to the fraction of the atmosphere, by number, made up a particular molecule. Therefore to recover the local water content of the atmosphere we would multiply the fractional abundance of water by the number density. \\
Initially, ablation leads to an order of magnitude increase in the mean water abundance in the very outer atmosphere (driven by a three orders of magnitude increase in abundance at the impact site), a relative enhancement which decreases as we move deeper into the atmosphere where the local density (and hence absolute water content) increases. However eventually we reach the point ($\sim5\times10^{-3}$ bar) where the ram pressure of the atmosphere exceeds the tensile strength of the cometary ice, leading to cometary breakup and a massive deposition of water which drives a two order of magnitude increase in the local fractional water abundance. Note that, even though a significant fraction of the cometary mass is deposited at pressures $>0.1$ bar (see \autoref{fig:energy_mass_deposition}), { the relative enhancement seen here is small due to the already high density of this region. }
Evidence for this weak enhancement of the deep atmosphere can also be seen in the Appendix (\autoref{fig:combined_time_evolution}), where we plot the variation of water abundance with time.    \\
The behaviour of the water associated with these two different sources of deposition differs as our model evolves. In the outer atmosphere, the ablated water rapidly settles and freezes out of the atmosphere, leading to an abundance profile which is only weakly enhanced with respect to our reference state around two-months post-impact. 
On the other hand, the water which is deposited deeper into the atmosphere due to the cometary breakup starts to slowly mix vertically, leading to a water enrichment in the mid-atmosphere that slowly shifts to lower pressures. With time, this vertical mixing carries water aloft into the outer atmosphere, leading to a fractional water abundance profile which is again significantly enriched, by over an order of magnitude,  with respect to our reference state. \\
This vertical mixing driven enhancement of the fractional water abundance in the outer atmosphere is most apparent in the multi-year evolution profiles plotted on the right-hand-side of \autoref{fig:combined_curves_water_temp}. 
We see that the fractional water abundance at all pressures $\lesssim5\times10^{-3}$ bar peaks approximately two years post-impact, after which time water slowly settles/freezes out of the atmosphere. Eventually, we reach the point, around fifteen to nineteen years post-impact, that the overall fractional water abundance is close to that found in our reference state, finding a slight, approximately $10\%$, enhancement in the outer and mid-atmosphere that persists to quasi-steady-state. \\

Moving onto the effect of the cometary impact on the temperature of our exo-Earth analogue's atmosphere, we find something rather distinct from the mid-atmosphere heating found in the tidally-locked impact model of \citetalias{2024arXiv240911151S}: bar a few exceptions (discussed below), we find a near-uniform cooling of the atmosphere between $0.1$ and $\sim5\times10^{-7}$ bar, as shown on both the bottom row of \autoref{fig:combined_curves_water_temp} and in the time evolution of the average temperature ( see \autoref{fig:combined_time_evolution} in \autoref{sec:appendix}). \\
The exceptions to this near uniform cooling are driven by two effects. The first is that the cometary impact delivers kinetic energy to the outer and mid atmosphere driving short-lived heating that can be seen in the one month post-impact profile. Whilst the second is driven by the seasonality of our model. Our exo-Earth analogue's orbit includes an Earth-like axial tilt (i.e., obliquity) which means that the relative insolation of the northern and southern hemispheres also varies, leading to differences in the mean temperature of the atmosphere due to differences in the surface land-ocean coverage.
Thus to isolate the effects of the impact alone on the temperature of the atmosphere, we must look at profiles calculated using averages of the same month (bottom right of \autoref{fig:combined_curves_water_temp}). Here we find a consistent cooling which peaks in strength when the water content of the mid and outer atmosphere is at its highest. \\
But what exactly is driving this cooling? And why do we find that the atmosphere is coolest when the water abundance is at its highest when \citetalias{2024arXiv240911151S} found the reverse? 
The heating found by \citetalias{2024arXiv240911151S} occurs due to strong irradiation of water at the sub-stellar point, with the optically-thick water absorbing a significant fraction of the incoming energy.
The inclusion of a diurnal cycle and the lack of a dayside upwelling here (see \autoref{sub:advection_of_water}) means that this heating effect does not occur in our exo-Earth-model and instead we find both stratospheric and mesospheric cooling. \\
One possible driver of this cooling might be the reduction in atmospheric ozone which in turn reduces the absorption, by ozone, of UV irradiation. As we discuss in \autoref{sub:effects_of_water_on_composition}, the influx of water and its eventual photodissociation drives a significant (order of magnitude) reduction in atmosphere ozone abundance between $10^{-3}$ and $10^{-7}$ bar (see \autoref{fig:H_OH_HOx_NOx_H2_O3_evo}), a pressure range which corresponds to the peak in cooling found in our model. Further, much like the water content of the mid and outer atmosphere, this ozone depletion is also strongest two years post-impact. \\
Speaking of water, it is possible that the impact driven enhancement in water vapour between the tropopause and stratopause may strengthen the radiative cooling of the atmosphere \citep{1999GeoRL..26.3309D}, as was the case for the Earth in 2022 when an undersea volcanic eruption lead to a massive upwelling of water vapour and localised cooling \citep{Stocker2024}.   \\
Finally, the formation of clouds and cloud ice can also affect the temperature of the atmosphere by changing the albedo of the atmosphere, and hence, the total energy absorbed by the planet. For example, the formation of clouds at the impact site (see \autoref{fig:cloud_fraction} in \autoref{sec:appendix}) might lead to a local enhancement in albedo and hence cooling. However our analysis of the evolution of the mean cloud fraction, as well as the longwave and shortwave cloud radiative forcing (i.e., the influence of clouds on our exo-Earth analogue's radiative budget), suggests that effects associated with clouds are short-lived. On the other hand, the delayed advection of water into the outer atmosphere also delays the formation of low pressure cloud ice (see \autoref{fig:cloud_ice} in \autoref{sec:appendix}), another source of albedo (via scattering) which peaks two years post-impact, as can be seen in our synthetic reflected light spectra. Note that such an effect may also have occurred in the tidally-locked models of \citetalias{2024arXiv240911151S}, but was masked by the fixed insolation of the dayside and the associated heating by optically thick water vapour. 

\begin{figure*}[tbp]
\begin{centering}
\includegraphics[width=0.95\textwidth]{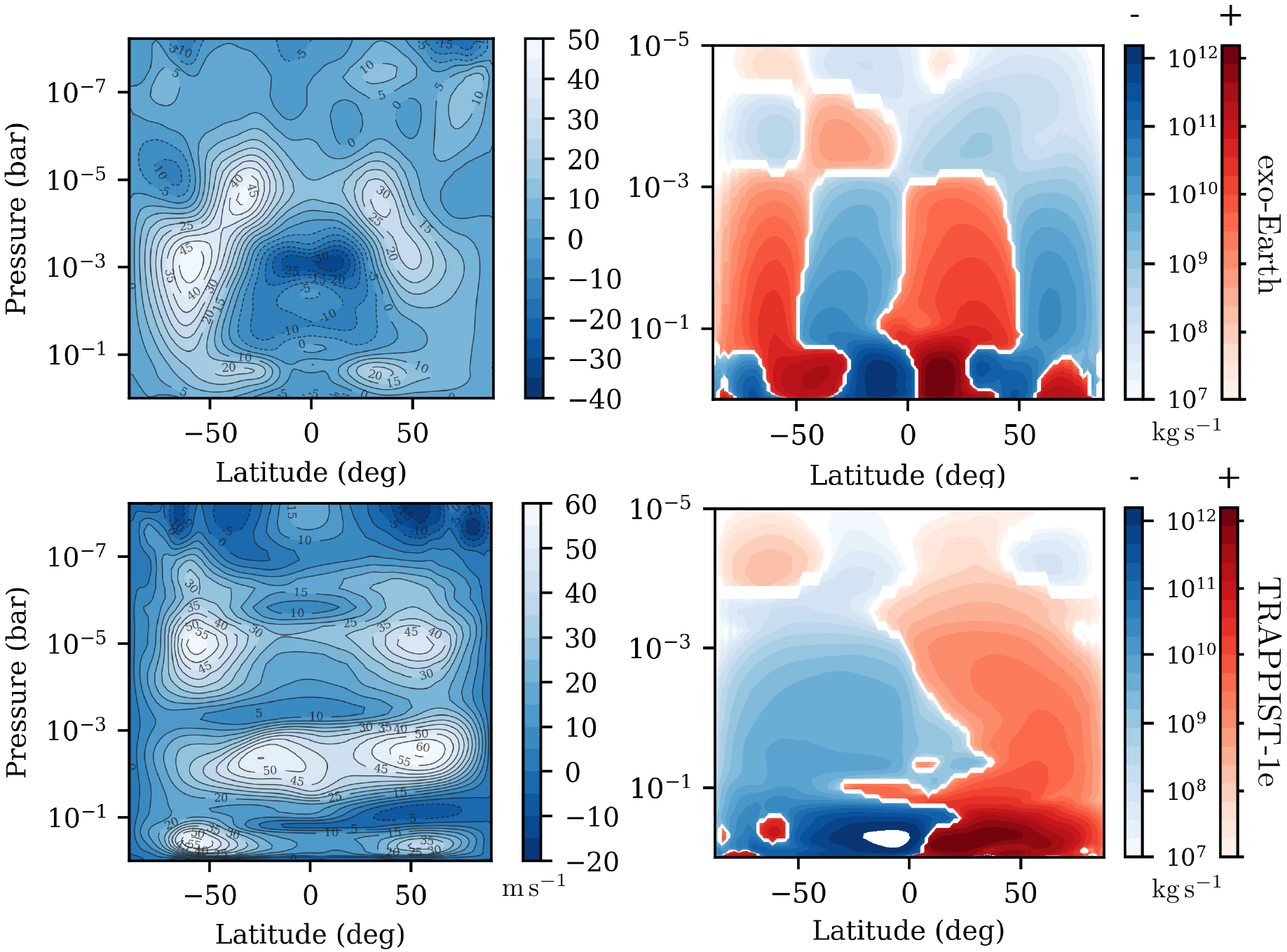}
\caption{ Zonally and temporally averaged zonal-wind (left) and meridional circulation streamfunction (right) profiles for both our exo-Earth analogue model (top) and the tidally-locked, TRAPPIST-1e-like, model (bottom) of \citetalias{2024arXiv240911151S}. Note that the meridional circulation profile is plotted on a log-scale with clockwise/anti-clockwise circulations shown in red/blue respectively. We do not plot the meridional circulation for $P<10^{-5}$ bar due to the relative weakness of outer atmosphere (i.e. very low pressure) circulations.  \label{fig:zonal_wind_streamfunction_extra} }
\end{centering}
\end{figure*}

\begin{figure*}[tbp]
\begin{centering}
\includegraphics[width=0.95\textwidth]{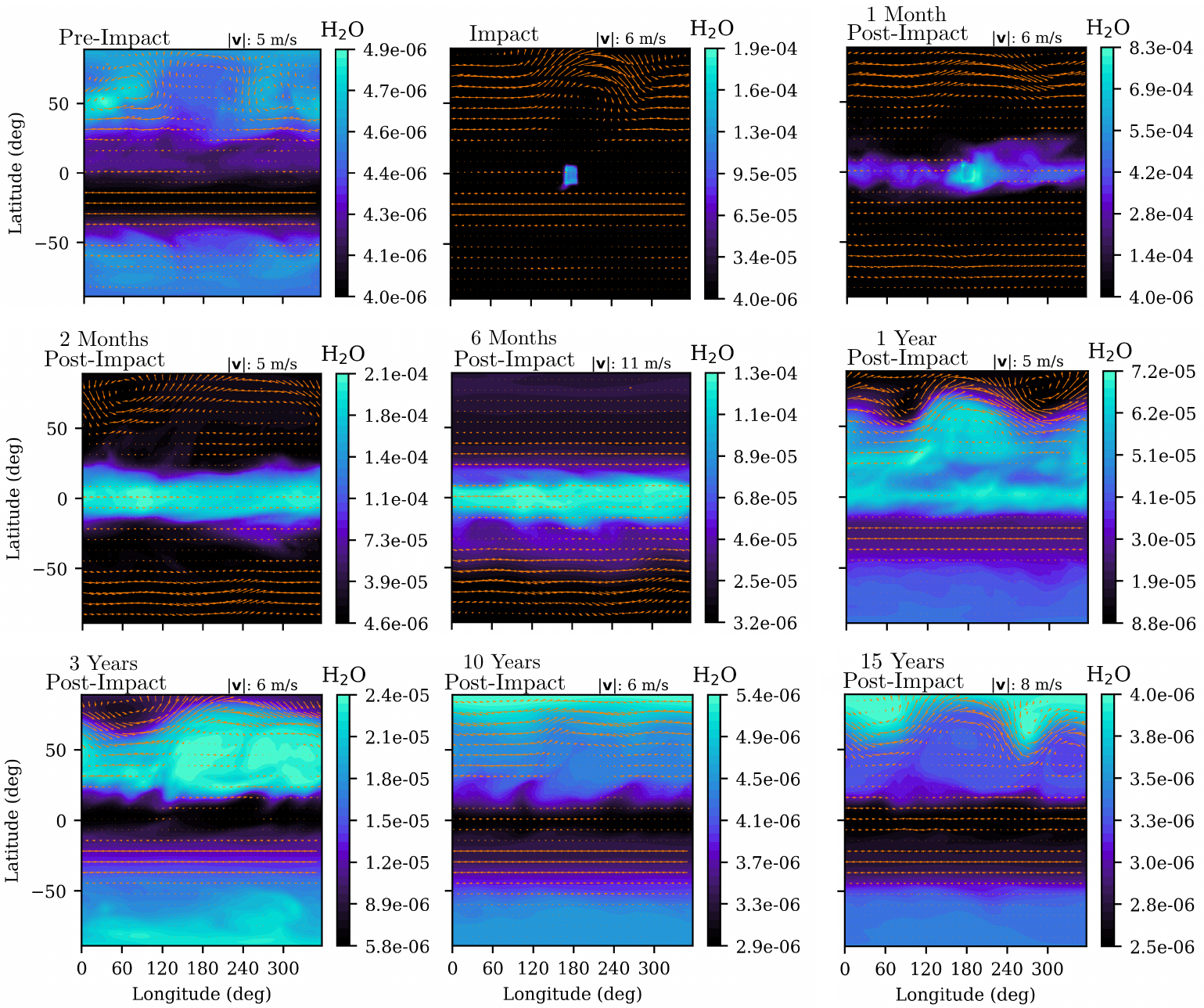}
\caption{ Horizontal slices of the fractional water abundance at a pressure of $2\times10^{-2}$ bar (corresponding to an altitude of $\sim24\,$km), showing the transport of cometary water from the impact site, over the pacific ocean at the equator, towards the poles. This transport is closely associated with the horizontal wind, the mean ($|\bm{v}|$) of which is shown at the top right of each panel, and which we plot using orange quivers. Note that the dynamic range of each colour bar is different in order to highlight the change in water distribution with time and that each of the annual snapshots is calculated by averaging over the month of February.   \label{fig:water_evolution} }
\end{centering}
\end{figure*}

\subsection{Advection of Water} \label{sub:advection_of_water}
When comparing our exo-Earth analogue impact model with the tidally-locked, TRAPPIST-1e-like, impact model of \citetalias{2024arXiv240911151S}, whilst the breakup altitudes are broadly similar (with the comet breaking up $1$ km deeper for our exo-Earth analogue), the resulting enhancement in atmospheric water vapour is very different, particularly at low pressures. To understand why this is the case, we next look at the atmospheric circulations, and how they are influenced by the inclusion of a diurnal cycle, with a particular focus on the vertical transport of water. \\

In \autoref{fig:zonal_wind_streamfunction_extra}, we plot the zonally-averaged zonal-wind (left) and meridional circulation streamfunction (right) for both our exo-Earth analogue (top) and the tidally-locked atmosphere of \citetalias{2024arXiv240911151S} (bottom). \\
The zonal-mean zonal-wind shows evidence of a significant north-south asymmetry, particularly at the pressure range where we find that the deposited cometary water has a maximal, initial, effect on the fractional water abundance (i.e, between $10^{-4}$ bar and $10^{-1}$ bar). Here we find a jet in the southern hemisphere but not in the northern hemisphere.
The underlying cause of this is the orography of the Earth. In the southern hemisphere not only is there significantly less land, but there is also a band of latitudes, centred at $\sim-60^{\circ}$ (i.e. the centre of our $10^{-3}$ bar jet), that are almost entirely land free (i.e, there are almost no obstacles to zonal wind flow).
A similar orographic effect also occurs in the tidally-locked models of \citetalias{2024arXiv240911151S}, but here, differences in the global circulation regime lead to differences in the expression of this asymmetry. For example, \citetalias{2024arXiv240911151S} find that the winds between $0.1$ and $10^{-3}$ bar are stronger in the northern hemisphere, with a stronger southern hemisphere jet forming near the surface, again in this land-free band at $\sim-60^{\circ}$ latitude. \\
But what is different about the global circulation regime of a Earth-like planet with a diurnal cycle and one which is tidally locked? And why do these differences occur? Very briefly, there are two primary drivers of {\it large-scale} winds in planetary atmospheres, stellar insolation which drives differential heating and hence thermal/pressure gradients, and rotation, which shapes atmospheric circulations via Coriolis forces (see \citealt{2019ApJ...871..245K} for a discussion of these effects on global circulations). For the two regimes we discuss here, it is primarily differences in the insolation pattern that primarily drive differences in the atmospheric circulations. On the Earth, the primary driver of atmospheric circulations is differential heating between the equator and the poles, whereas on a tidally locked planet it is instead differential heating between the permanently illuminated dayside and the cold, dark, nightside. Furthermore, on a tidally-locked planet, the static nature of this insolation with respect to the planetary surface leads to the formation of global-scale standing wave structures which drive planetary scale motions.
This leads to differences in the wind structure both zonally, as discussed above, but also latitudinally (i.e., between the equator and poles). \\

To investigate the latitudinal winds, and their associated material transport, we next look at the meridional circulation streamfunction $\psi$, which takes the form,
\begin{equation}
	\psi = \frac{2\pi R_{p}}{g\cos\theta}\int^{P_0}_{P_{top}}vdP,
\end{equation}
where $v$ is the latitudinal velocity, $R_{p}$ is the radius of the planet, $g$ is the surface gravity, $\theta$ is the latitude, and $P_{0}$/$P_{top}$ are the pressure at the top/surface of the atmosphere respectively. The right-hand side of \autoref{fig:zonal_wind_streamfunction_extra} shows the meridional circulation profile for our exo-Earth analogue (top) and the tidally-locked atmosphere of \citetalias{2024arXiv240911151S} (bottom). Here clockwise circulations are show in red whilst anti-clockwise circulations are shown blue. Where these circulations meet, net flows develop. For example, both models reveal a net upflow near the equator, although this upflow does not extend to as high altitudes in our exo-Earth analogue as it does for the Earth-like TRAPPIST-1e. At its core, this difference in both the strength and extent of the equatorial upwelling occurs because of differences in the circulation regimes of our two models, with a weakening of vertical mass transport in our exo-Earth analogue explaining the slower/weaker outer atmosphere fractional water abundance enhancement found here (see \autoref{sub:water_abundance_and_mean_temperature}) when compared with \citetalias{2024arXiv240911151S}. \\
On the Earth, and in our exo-Earth analogue, the meridional circulation profile in each hemisphere consists of three circulation cells. Near the equator we have the Hadley cells, which consist of an upflow at the equator and a downflow at mid-latitudes, in the so-called inter-tropical-convergence zone. Note that whilst this mass flow does extend all the way to the stratopause, it is strongest in the troposphere (i.e., for $P>0.1$ bar). It becomes significantly weaker in the stratosphere due to the increase in temperature with altitude leading to warmer layers lying above cooler layers, a structure which reduces the strength of convective, and turbulent, mixing.  
Moving to higher latitudes, we find the Ferrel cells, with a downflow in the inter-tropical-convergence zone and an upflow at high latitudes/over the poles. Finally, in the troposphere, we find the polar cells which are associated with the downwards transport of cold air over the poles. Whilst, as we discuss below, this multi-celled circulation structure is highly efficient at mixing material latitudinally, its ability to carry material aloft into the low-pressure regions of the atmosphere is significantly weaker than the circulation found in a tidally-locked atmosphere. On these tidally-locked worlds, the strong day-night irradiation contrast drives an upwelling on the dayside which is balanced by a downwelling on the nightside, the so-called global overturning circulation (for more details see, for example, \citealt{2021PNAS..11822705H}). In the meridional circulation profile, this can be seen as a strong upwelling near the equator which extends from the surface to $P\simeq10^{-5}$ bar, and, as was seen in Figure 2 of \citetalias{2024arXiv240911151S}, is particularly efficient at carrying water aloft. \\

To investigate the strength of this multi-cell mediated latitudinal mixing, we next explore the horizontal water vapour transport at a pressure of $2\times10^{-2}$ bar. \autoref{fig:water_evolution} shows the fractional water abundance at this pressure at nine different points in time, ranging from pre-impact (top left) to fifteen years post-impact (bottom right). Note that, beyond this point the evolution to quasi-steady-state  is slow, with differences in the profiles at $2\times10^{-2}$ bar being primarily associated with multi-year oscillation cycles.  \\
Initially water is deposited at the equator over the pacific ocean (top middle), and from here equatorial winds mix this water (top right) to form a mostly homogenous band of water at the equator two months post-impact (middle left). Next, the water slowly disperses latitudinally (centre), likely due to the poleward component of the Hadley cells, until around a year post-impact (middle right) at which point in time we find that the peak in water abundance has shifted to mid-latitudes in the northern hemisphere. Here, the water abundance profile is highly shaped by the off-equator winds, and with a significant fraction of the impact delivery water becoming trapped in a vortex centred over the northern pacific ocean. At the same time, the vortex over the north-pole keeps the impact delivered water away from this region, even as water is advected to the south pole (bottom left). 
However, eventually, variations in the wind associated with the long time-scale oscillation cycles lead to a weakening of this polar vortex (bottom middle) and water can accumulate at high latitudes, over the poles (bottom right), at least when seasonally appropriate. For the layer of the atmosphere considered here (with an altitude of $24$ km), in both our impacted and reference models, we find that the fractional water abundance over each pole peaks in winter, hence the peak in water abundance at the north pole in February shown in the bottom right panel of \autoref{fig:water_evolution}. \\
Comparing \autoref{fig:water_evolution} with figure 4 of \citetalias{2024arXiv240911151S}, we can see a difference in the both the form and strength of the latitudinal transport, both of which can be linked back to the meridional circulation profile. For our exo-Earth analogue, the meridional circulation profile is generally symmetric about the equator, and this leads to an initially symmetric latitudinal dispersion of the deposited water vapour. However the symmetry is not perfect and at $\sim10^{-2}$ bar, equatorial regions are slightly more associated with the northern hemisphere's Hadley cell, leading to the asymmetry found one year post-impact. The multi-cell circulations combined with the seasonality of our model also drive the breakup of zonal bands of water much faster than in the tidally-locked scenario. \\
Overall our results suggest that, when compared with an almost equivalent tidally-locked counterpart (\citetalias{2024arXiv240911151S}), horizontal mixing of impact delivered water is more efficient in our exo-Earth analogue, whilst vertical mixing and transport to high altitudes is significantly less efficient. As we discuss in \autoref{sub:observational_implications}, this may have significant implications for the observability of individual, massive, cometary impacts. 

\begin{figure*}[tbp]
\begin{centering}
\includegraphics[width=0.776\textwidth]{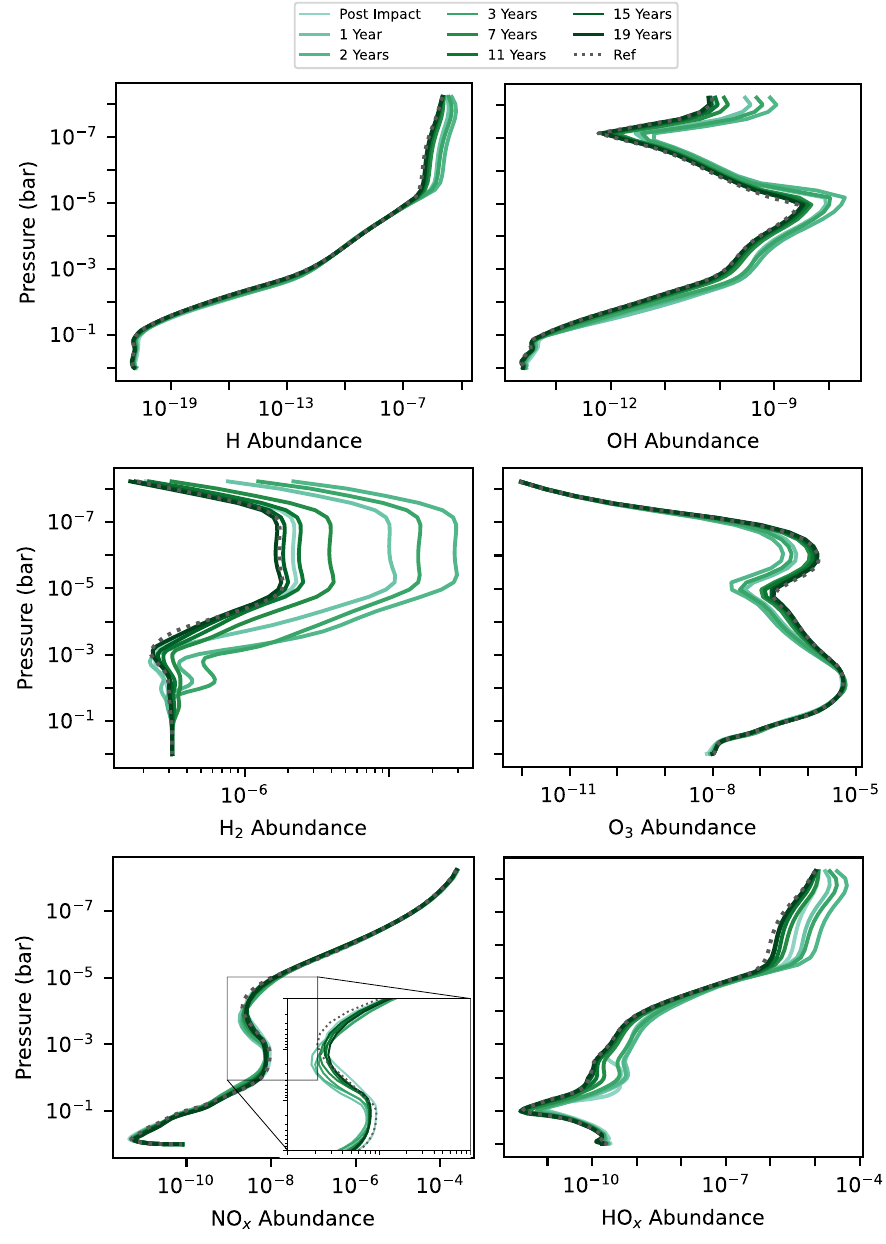}
\caption{ Fractional atomic hydrogen (\ce{H} - top left), the hydroxyl radical (\ce{OH} - top right), molecular hydrogen (\ce{H2} - middle left), ozone (\ce{O3} - middle right), \ce{NOx} (bottom left), and \ce{HOx} (bottom right) abundance profiles selected to exemplify the post-impact changes in atmospheric composition and chemistry over the approximately twenty years required for the atmosphere to settle into a quasi-steady-state reminiscent of our reference atmosphere (grey-dashed).
Each profile is calculated by averaging both horizontally over all latitudes and longitudes, and temporally over the month of February for seven different points in time post-impact. To better demonstrate the impact-induced changes in \ce{NOx} chemistry, a zoomed-in view of the \ce{NOx} abundance profile between $10^{-5}<P<10^{-2}$ bar has been included.      \label{fig:H_OH_HOx_NOx_H2_O3_evo} }
\end{centering}
\end{figure*}
\subsection{Effects of Water on Composition} \label{sub:effects_of_water_on_composition}
Before we explore the observational implications of this weakened vertical mixing, we next investigate how the influx of water and thermal energy associated with the cometary impact affect the overall composition of our exo-Earth analogue's atmosphere. \autoref{fig:H_OH_HOx_NOx_H2_O3_evo} shows six molecules which exhibit a significant change in abundance due to the influx of cometary water at seven points in time between the cometary impact and the model approaching a quasi-steady-state.    \\

We start by exploring the direct products of water photolysis, which is one of the main mechanisms via which both hydrogen and oxygen can be released from the impact delivered water and hence, go on to affect the broader chemistry. The primary water photodissocation pathway in our exo-Earth analogue is:
\begin{equation}
	\ce{H2O} + \ce{$h$\nu} \rightarrow \ce{H} + \ce{OH},
\end{equation}
leading to the formation of atomic hydrogen (\ce{H} - top left) and the hydroxyl radical (\ce{OH} - top right). To a lesser extent, with a peak reaction rate of around a fifth of the above reaction rate at $10^{-5}$ bar, we also find that the photodissocation of water can also lead to the formation of molecular hydrogen (\ce{H2} - middle left) and excited oxygen (\ce{O(1D)} - not shown here), although the short lifetime of the latter, and its destruction by water ($\ce{O(1D)} + \ce{H2O} \rightarrow 2\,\ce{OH}$), means that `large' changes in \ce{O(1D)} abundance are short-lived,
\begin{equation}
	\ce{H2O} + \ce{$h$\nu} \rightarrow \ce{H2} +\ce{O(1D)}. 
\end{equation}
The formation of both hydrogen (\ce{H} and \ce{H2}) and the hydroxyl radical (\ce{OH}) are significantly delayed with respect to the cometary impact. For all three species the abundance peaks approximately two years post-impact, around the same time that the water abundance for pressures $<10^{-3}$ bar, also peaks (see \autoref{fig:combined_curves_water_temp}). \\
As the simulation progresses towards quasi-steady-state, the abundance of \ce{H} and \ce{OH} decreases faster than the abundance of water. This occurs because, as the abundance of water in the outer and mid atmosphere decreases, so too does the rate of photodissociation and hence \ce{H}, \ce{H2}, and \ce{OH} formation. This decrease in the formation rate then leads to a net decrease in abundance as the enriched species are rapidly consumed by further reactions linked with the atmosphere evolving towards an equilibrium state. \\
Most prominently, the hydroxyl radical can react with, and destroy, ozone (\ce{O3}) leading to the formation of the hydroperoxyl radical (\ce{HO2}),
\begin{equation}
	\ce{OH} + \ce{O3} \rightarrow \ce{HO2} + \ce{O2}.
\end{equation}
This enhancement of \ce{HO2} and destruction of \ce{O3} can clearly be seen in \autoref{fig:H_OH_HOx_NOx_H2_O3_evo}, where we plot both the time-dependent \ce{O3} profile (middle right) and \ce{HOx} profile (bottom right). Note that \ce{HOx} simply represents the sum of the abundances of \ce{H}, \ce{OH}, \ce{HO2} and \ce{H2O2} with its change in abundance primarily reflecting changes in both \ce{OH} and \ce{HO2} abundance in our impacted atmosphere.  \\
The hydroperoxyl radical, and \ce{HOx} family molecules more generally, play a key role in the destruction of ozone in the Earth's atmosphere, with \ce{HO2} alone account for around half of the ozone destruction on Earth \citep{doi:10.1126/science.266.5184.398}. Destruction of \ce{O3} by \ce{HO2} also leads to the formation of \ce{OH},
\begin{equation}
 	\ce{O3} + \ce{HO2} \rightarrow \ce{OH} + \ce{2O2}, 
 \end{equation} 
thus forming a catalytic cycle of ozone destruction which takes the form,
\begin{equation}
	\ce{OH} + \ce{2O3} \rightarrow \ce{OH} + \ce{3O2}.
\end{equation}
This cycle explains why we find such a strong decrease in ozone abundance post impact. The persistence of \ce{OH}/\ce{HO2} (i.e., \ce{HOx}) enrichment to quasi-steady-state helps to explain why we also find a long-lasting decrease in ozone abundance at low pressures. However, note that, due to this destruction taking place at low pressures, the overall changes to the total ozone column density are small, on the order of a few Dobson Units (one Dobson Unit is $2.69\times10^{20}\,\mathrm{mol\,m^{-2}}$), which represents approximately a $\sim1\%$ change in the global average ozone column density with respect to the Earth ($\sim320$ DU in our reference atmosphere) at quasi-steady-state. However small differences can add up, and it is possible that a multi-impact or continuous bombardment scenario may drive larger changes in the atmospheric ozone abundance. Since ozone plays an important role in shielding the surface from harmful UV radiation, this may have important implications for planetary habitability (see, for example, \citealt{2006IJAsB...5..295G} for a discussion of the role that ozone may have played in the early Earth's habitability). \\
Moving onto \ce{NOx} (the sum of \ce{N}, \ce{NO}, and \ce{NO2} abundances), we find a profile that is primarily dominated by the abundance of \ce{NO}. In general, we find that the changes to the abundance of \ce{NO} (and other \ce{NOx} family molecules) are small. They include an slight increase in abundance for pressures $\gtrsim5\times10^{-2}$ bar, likely due to the post-impact oxygenation of the atmosphere, and a slight shift in the shape of the profile between $10^{-5}$ and $10^{-2}$ bar which we attribute to a combination of factors, including stratospheric/mesospheric cooling and the destruction of low pressure ozone. \\ 

Overall, we find that the additional hydrogen and oxygen which the comet delivers drives small, but persistent, changes to the composition of the planet. This includes a slight enrichment of, for example, \ce{H2O}, \ce{OH}, { and} \ce{O2}, and a slight depletion of \ce{O3}. These long-lasting changes suggest that cometary impacts, and in particular ongoing bombardment or massive impacts, have the ability to drive persistent changes in the atmospheric compositions of Earth-analogue exoplanets.

\begin{figure*}[tbp]
\begin{centering}
\includegraphics[width=0.85\textwidth]{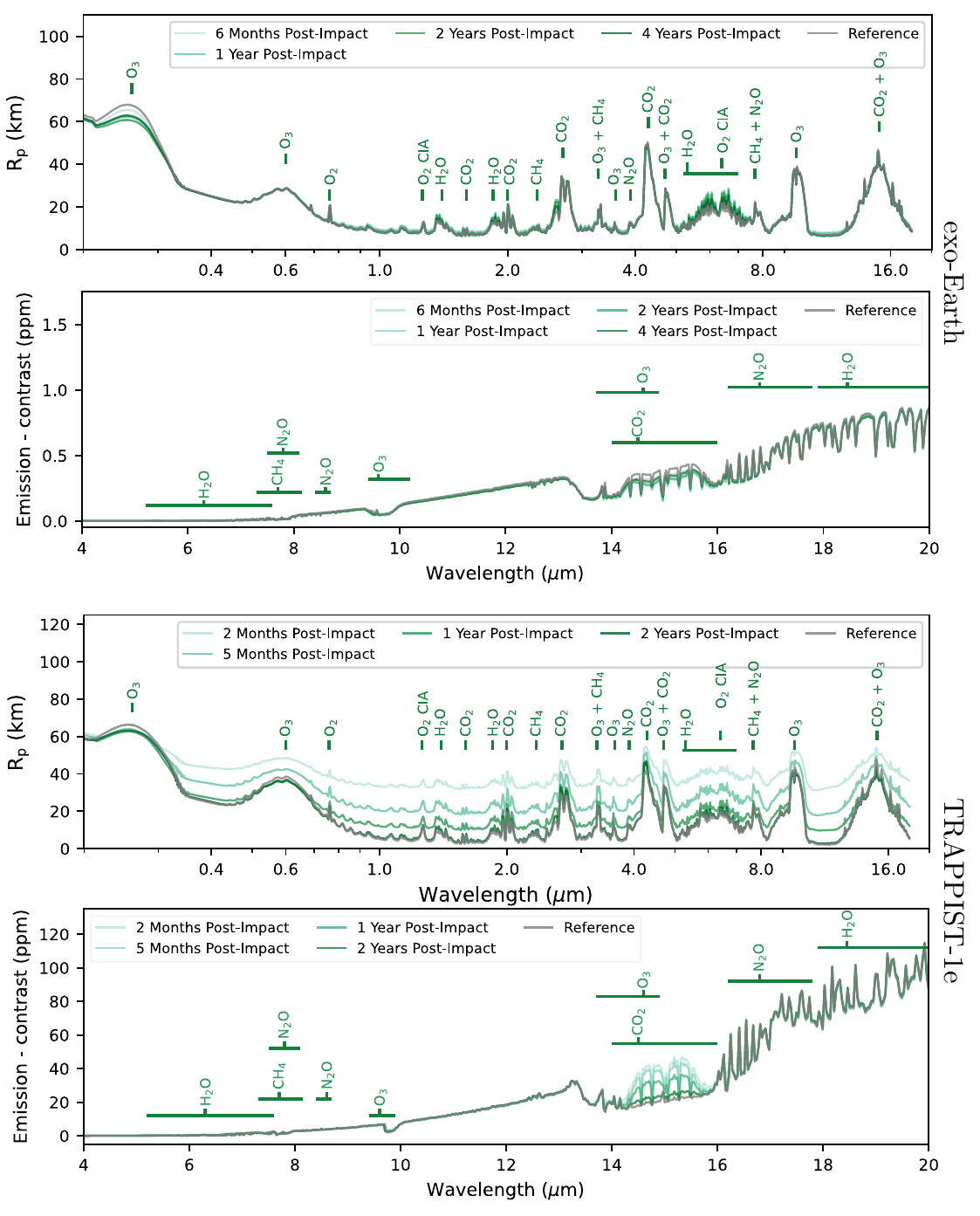}
\caption{ Synthetic transmission (1st and 3rd rows, in units of transit atmospheric thickness; R$_p$), and emission (2nd and 4th rows, in units of contrast; radiance ratio with respect to the host star in parts per million) spectra for both our exo-Earth analogue model (top group), and the tidally-locked model of \citetalias{2024arXiv240911151S} (bottom group). In addition to reference spectra calculated from our reference state (grey), we plot transmission and thermal emission spectra at four points in time post-impact: 6 months, 1 year, 2 years, and 4 years. 
To aid in interpretation, we have labelled a number of spectroscopic features of interest, including \ce{H2O}, \ce{CO2}, \ce{CH4}, \ce{N2O}, \ce{O2}, and \ce{O3}.  \label{fig:combined_spectra} }
\end{centering}
\end{figure*}
\subsection{Observational Implications} \label{sub:observational_implications}
\begin{figure*}[tb]
\begin{centering}
\includegraphics[width=0.85\textwidth]{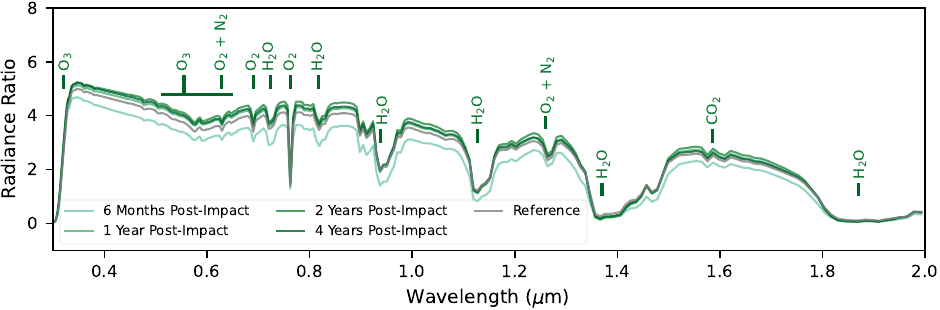}
\caption{Synthetic reflection spectra, in units of contrast (radiance ratio of reflected light to host star emission) for our exo-Earth analogue cometary impact model. In addition to reference spectra calculated from our reference atmosphere (grey), we plot reflection spectra at four points in time post-impact: 6 months, 1 year, 2 years, and 4 years.  To aid in interpretation, we have labelled a number of spectroscopic features of interest, including \ce{O3}, \ce{O2}, \ce{H2O}, \ce{CO2} and \ce{N2}. \label{fig:reflection_spectra}}
\end{centering}
\end{figure*}

We finish by investigating if impact-induced changes in the atmospheric chemistry, composition, or climate of our exo-Earth analogue might be observable with either current (e.g., JWST) or future (e.g., the Habitable Worlds Observatory or the ELT) telescopes. \\
We use the Planetary Spectrum Generator (PSG; \citealt{2018JQSRT.217...86V}) to calculate idealised transmission spectra spanning the UV, optical, and near- to mid-infrared ($0.2$ to $20\,\mathrm{\mu m}$; \autoref{fig:combined_spectra}), thermal emission spectra in the mid-infrared ($4$ to $20\,\mathrm{\mu m}$; \autoref{fig:combined_spectra}), and reflected light spectra in the near-uv, optical, and near-infrared ($0.3$ to $2\,\mathrm{\mu m}$ - \autoref{fig:reflection_spectra}).  For all three spectra, we consider our exo-Earth analogue planet to be in an Earth-like orbit around a sun-like star 10 parsecs away from the observer. When simulating both the thermal emission and reflected light spectra for our exo-Earth analogue, we assume that the observing telescope includes a coronograph in order to mask the luminous host star and maximise the light captured from the planet. 
The emission and transmission spectra (shown in \autoref{fig:combined_spectra}) are calculated with a spectral resolution of $R=250$, whilst the reflected light spectra (shown in \autoref{fig:reflection_spectra}) has a reduced resolution of $R=140$. All spectra are calculated using 3D mapping for the observed surface or terminator where appropriate and we consider four points in time post-impact (6 months, 1 year, 2 years, and 4 years) in order to demonstrate the effect that vertical mixing has on observations. For this same reason, and to aid in comparisons, in \autoref{fig:combined_spectra}, we also reproduce the synthetic transmission and thermal emission spectra from the tidally-locked model of \citetalias{2024arXiv240911151S} (their figure 11).   \\

We start by exploring the effects of the cometary impact on the apparent radius of our exo-Earth analogue during transit, as shown on the top row of \autoref{fig:combined_spectra}. It is apparent that our icy cometary impact has a minimal (observable) effect on the pressures probed by transmission spectroscopy. 
The primary changes we observe are a general strengthening of any water features, for example the narrower $1.9\,\mathrm{\mu m}$ feature or the broader feature between $5.2$ and $7\,\mathrm{\mu m}$, and a weakening of ozone features, in particular the short wavelength ozone feature at $0.255\,\mathrm{\mu m}$ (i.e., the Hartley band). However, unlike \citetalias{2024arXiv240911151S} (row three of \autoref{fig:combined_spectra}) we do not find a significant, observable, increase in the `continuum' level of our transit spectrum. 
This can be linked back to the inefficient vertical advection of water (\autoref{sub:water_abundance_and_mean_temperature}) combined with the deep break-up site.
Together this leads to a muted enhancement in low-pressure cloud ice (see \autoref{fig:cloud_ice} in \autoref{sec:appendix}) when compared with the tidally-locked case in which low pressure cloud ice is massively enriched (from near zero to making up a few tenths of a percent of the atmosphere) by the vertical transport of water (\autoref{fig:zonal_wind_streamfunction_extra}) associated with the global overturning circulation. 
Note that scattering does slightly increase the observed `continuum' level, however these changes, as well as the changes to the strength of the water and ozone features, would be difficult if not impossible to detect with JWST, especially when considering the difficulty of capturing (and stacking) multiple observations of a planet with an orbital period of a year. \\
Evidence for the delayed vertical advection of water can be seen in the $0.255\,\mathrm{\mu m}$ ozone feature. Here we find that, six months post-impact, the apparent radius (i.e., strength of the ozone absorption) of our exo-Earth analogue has decreased from $68.0$ km to $65.5$ km. Then, as the advection of water from the mid-atmosphere starts to grow in magnitude and the water photolysis rate rises, leading to the destruction of low pressure ozone (\autoref{sub:effects_of_water_on_composition}), we find that the altitude probed by the transit spectra decreases to $62.5$ km one year post-impact and $61.0$ km two years post-impact, when the mid-atmosphere ozone abundance is at its minimum. Finally, as water transport, and hence, photolysis slows and the ozone abundance of the outer atmosphere starts to recover, we find that the altitude probed increases to $63$ km four years post-impact. However probing this change observationally will be challenging as it falls into the ultraviolet, with no exoplanet-focused (space) missions planned to observe at these wavelengths until, potentially, the Habitable Worlds Observatory\footnote{\url{https://habitableworldsobservatory.org/}}.  \\

Moving on to the thermal emission spectra, shown on the second row of \autoref{fig:combined_spectra}, we find a clear increase in the strength of the \ce{CO2} absorption feature between $\sim14$ and $\sim16\,\mathrm{\mu m}$. This can likely be attributed to the post-impact cooling of the stratosphere. As discussed in \citet{2000ESASP.451..133S}\footnote{see also \citet{2018ApJ...854...19R}}, the strength and shape of the \ce{CO2} absorption feature around $15\,\mathrm{\mu m}$ is linked to both the temperature and temperature gradient of the stratosphere, which itself is set by the ozone abundance (via local heating - \autoref{sub:water_abundance_and_mean_temperature}). As such, this \ce{CO2} feature is essentially an indirect signature of ozone, which also explains why the change in feature strength is temporally aligned with the destruction of ozone. \\
The response of this $\sim14$ to $\sim16\,\mathrm{\mu m}$ \ce{CO2} feature to the influx of material associated with the cometary impact is also rather different in the tidally-locked atmosphere of \citetalias{2024arXiv240911151S}. There, the strong vertical advection of water to the outer atmosphere, as well as the formation of low-pressure cloud ice, essentially acts to mask the deeper \ce{CO2} feature, and hence the emission between $\sim14$ to $\sim16\,\mathrm{\mu m}$ is closer to a black-body (see the bottom row of \autoref{fig:combined_spectra}), increasing the apparent emission at these wavelengths. 
Additionally, the shape of the feature at $15\,\mathrm{\mu m}$ is different. This can be linked to the water opacity driven heating of the mid atmosphere as the $15\,\mathrm{\mu m}$ \ce{CO2} feature is known to be sensitive to both the stratospheric temperature and temperature gradient \citep{2000ESASP.451..133S}.\\
However the biggest difference is the strength of the thermal emission relative to the host star, i.e., the contrast. For the tidally-locked world of \citetalias{2024arXiv240911151S}, the coolness of the host M-star (TRAPPIST-1) means that the thermal emission contrast, as well as the changes in strength of the \ce{CO2} feature, are on the order of tens to a hundred parts per million. On the other hand, the increased luminosity of a Sun-like star compared with a cool M-dwarf ($L=0.0005L_\sun$; \citealt{2024ApJ...970L...4D}) means that thermal emission is much harder to distinguish, leading to a contrast ratio on the order of $0.1$ to one part per million and hence, changes in feature strength that are up to 100 times weaker than a similar planet orbiting a cooler star. \\

However, it is possible that the impact-induced changes to our exo-Earth analogue's climate and composition might be observed by measuring the reflection of visible and UV light off of the planetary atmosphere: such observations are referred to as reflected light spectra.  They rely on a combination of changes to the planets composition and climate modifying the strength of absorption features, particularly water absorption features at visible wavelengths, and changes to the albedo of the planet due to, for example, cloud formation or scattering (see, for example, \citealt{2022MNRAS.516.3167L}). \\

\autoref{fig:reflection_spectra} shows reflected light spectra for our exo-Earth analogue at four points in time post-impact, as well as for our reference atmosphere. \\
Shortly after the impact, the overall flux reflected by the planet is reduced when compared with our reference spectra. The primary driver of this is the seasonality of our exo-Earth analogue atmosphere:  the global climate state is rather different in July than February due to the obliquity of the Earth and differences in land and ocean coverage between the northern and southern hemispheres.
These differences can drive significant differences in reflected light spectra, as discussed by \citet{10.1093/mnras/stac2604}. Note that we do not consider seasonality when setting the phase of our observations in PSG. All reflected light spectra are calculated at an orbital phase of $90^\circ$, with the insolation centred over the impact site, i.e., mid-day occurring over the pacific ocean.    \\
As the simulation progresses the impact-driven changes to reflected light spectra become more apparent. We find that the overall fraction of the incoming light reflected back towards the observer has increased in concert with an increase in the strength of water absorption features. Both of these changes can be linked with the vertical advection of water from the break-up site, a link which is only further reinforced by the changes to the `continuum' level and water absorption feature strength peaking two years post-impact, i.e, the planets reflectivity is highest when the outer atmosphere is most water enriched. \\
The change in the `continuum' level, and hence the overall albedo of the planet, can be linked to the formation of low-pressure cloud ice (see \autoref{fig:cloud_ice} in \autoref{sec:appendix}), which scatters incoming light towards the observer, increasing the apparent brightness of the planet. 
As for the water absorption features, such as those found at $\sim0.95$, $\sim1.1$, and $\sim1.45\,\mathrm{\mu m}$, we find that the increase in strength of the absorption somewhat balances the increase in the overall reflectivity of the planet such that the minima of the $\sim0.95$ and $\sim1.1\,\mathrm{\mu m}$ features remains essentially unchanged between one and four years post-impact, and the minima of the $\sim1.45\,\mathrm{\mu m}$ feature is actually slightly deeper during that same period. \\
Future high spectral resolution observatories, such the the ELT, may be able to detect this combination of changes, however, even then such changes may be difficult to attribute to a cometary impact due to the inherent seasonality of an Earth-analogue planet as well as the affects of longer timescale climate trends.

{
\subsection{Caveats and Future Considerations}
In this work, we have studied the effects of an pure-water-ice cometary impact on the atmosphere of an Earth-analogue, i.e., \ce{N2}-\ce{O2}-dominated, exoplanet. However observations of exoplanetary systems have revealed a significant diversity in terrestrial planetary atmospheres, and observations of comets in our own solar system \citep{2015AA...583A...1L} and ices in interstellar molecular clouds \citep{2023NatAs...7..431M} reveals a diverse range of, water dominated, volatile compositions. As such, we suggest that future studies should work to loosen these restrictions, exploring cometary impacts across a wider parameter space in both atmospheric and cometary composition. \\

In our model, we have adopted the default land-mass distribution and lower boundary conditions which are tailored for the Earth. It has been shown that, for both the Earth and tidally-locked exoplanets \citep{sainsbury2024b}, the presence of land-masses can significantly shape the near-surface winds, including breaking symmetries in a way which might lead to localised effects, such as the accumulation of ozone over the south pole \citep{anand2024}. On the Earth, this symmetry breaking is driven by the difference in land-fraction between the northern ($\sim$67\% of the Earths-land) and southern ($\sim$33\%) hemispheres.  
However, on terrestrial exoplanets, whilst the presence of land is thought to be a key requirement for prebiotic chemistry (see, for example, \citealt{2020SciA....6.3419S,doi:10.1089/ast.2019.2187}), the land-ocean distribution is likely to differ from that of the modern Earth. As such, efforts should be made to explore how different land-ocean distributions shape atmospheric circulations and hence, the atmospheric response to an icy cometary impact. The tools required to study such effects are being developed as part of the next-generation release of CESM.\\

Cometary impacts are more likely to have occurred earlier in a terrestrial planet's lifetime, before the formation of an evolved atmosphere, such as that considered here form. In the solar-system, this may have happened around the time that planetary migration led to significant scattering of small bodies \citep{2003EM&P...92...89I,2004AdSpR..33.1524I,2014Icar..239...74O,2023PhyU...66....2M}. As such, models which consider the Earth's climate history are needed, including studying $\ce{CO2}$-dominated or $\ce{O2}$-poor atmospheres reminiscent of the Archean and early Proterozoic periods respectively. Studies of the latter are on particular interest as a low initial oxygen abundance present an opportunity for single, massive, impacts, to significantly affect the oxygenation state of the atmosphere.  \\
Furthermore, if these studies consider very young planets/atmospheres, then the stellar evolution of the host star should also be taken into account. For example, when the sun was younger, whilst the overall solar luminosity was lower, potentially cooling the planet, UV emissions were significantly higher \citep{Ribas_2005}, increasing photolysis rates.\\

As for the composition of the impacting comet, whilst water is likely to make up a significant portion of its volatile inventory, other volatiles, such as $\ce{CO}$, $\ce{CO2}$, $\ce{CH4}$, and $\ce{NH3}$ are expected to be present (see, for example, the review by \citealt{2004come.book..391B}). Although some of these volatiles are photosensitive (e.g., $\ce{CO2}$ and $\ce{CH4}$), and some are also destroyed by the products of water photolysis (i.e., $\ce{CH4}$) itself, at least a fraction of these volatiles should survive and alter the atmospheric chemistry. In particular, $\ce{NH3}$ deposition may lead to enhanced \ce{NO} and \ce{NO2} formation. Additionally, comets are composed of both ices and dust. Cometary dust will have a different ablation profile compared to that of the cometary ices during a cometary impact, and ablated dust particles may act as an additional opacity source, driving local heating or surface cooling.
Preliminary testing of a coupled model which includes the effects of cometary dust ablation and deposition is underway and future studies should consider more physically motivated cometary compositions in order to further quantify the role of impacts in volatile delivery. \\
Finally, cometary impactors of different sizes should also be investigated as this may effect where material is deposited in the atmosphere. For example, the impact and fragmentation model of \citet{2025MNRAS.539..376A} suggests that, for a comet with a very low heat transfer coefficient, as cometary size decreases, an increased fraction of the comets mass is deposited at low altitudes. For a high enough impactor flux, this may this may lead to enough material being deposited at high altitudes to be potentially observable, even for an Earth-like planet with weak vertical mixing. 
}
\section{Concluding Remarks} \label{sec:concluding}

In this work, we have coupled the cometary impact and breakup model of \citetalias{2024ApJ...966...39S} and \citetalias{2024arXiv240911151S} with an Earth analogue, terrestrial, (exo)-planetary atmospheric model calculated using the Earth system model, WACCM6/CESM2. We used this model to simulate the impact of a pure-water-ice comet (with $R=2.5\,$km and $\rho=1\,\mathrm{g\, cm^{-3}}$) over the equator and Pacific Ocean for an Earth-analogue planet orbiting a Sun-like star. Our model resembles the coupled impact model of \citetalias{2024arXiv240911151S} in many ways, such as the inclusion of an Earth-like land-ocean configuration and an initial atmospheric composition based upon the Earth pre-industrialisation, but it differs in one major aspect. This is that our Earth-analogue includes both a diurnal cycle (i.e. a day-night cycle) and seasonality due to a non-zero obliquity, whereas the TRAPPIST-1e-like model of \citetalias{2024arXiv240911151S} was both tidally-locked (i.e., the same face of the planet is always illuminated) and had a zero obliquity. This leads to major differences in the global circulation between the two models, differences which shape almost every key takeaway result of our study, and which we list below.
\begin{itemize}
	\item Although thermal ablation drives an initial enhancement of the fractional water abundance at low pressures, this enrichment is relatively short-lived due to a combination of water freezing out of the atmosphere and photodissociation.
	\item Longer lasting enrichment of the fractional water abundance can be linked to the vertical advection of water from the deep ($>5\times10^{-3}$ bar) breakup location of the comet.
	\item This vertical advection is relatively slow, and so we find that the fractional water abundance for pressures $<5\times10^{-3}$ bar peaks two years post-impact. In the tidally-locked model of \citetalias{2024arXiv240911151S}, the outer atmosphere water abundance peaked within six months of the impact.
	\item For pressures $<10^{-2}$ bar, this two years post-impact enhancement is over an order of magnitude. In the mid-atmosphere, at least a factor of two enrichment persists for over ten years post impact
	\item Whilst significant, this enhancement remains notably weaker than the multi-order-of-magnitude enhancement found in the tidally-locked models of \citetalias{2024arXiv240911151S}.
	\item This difference in the strength of the enhancement, as well as the delay in the post-ablation water enrichment of the outer atmosphere, can be linked to the global circulation pattern. In a tidally-locked atmosphere, the global overturning circulation associated with a permanent day-night contrast drives significant vertical advection of water. Whereas the multi-celled (Hadley, Ferrel and polar) circulation structure in our exo-Earth analogue is much less efficient at carrying water aloft, although it is more efficient at mixing the atmosphere latitudinally.
	\item Post-impact we find that both the stratosphere and mesosphere have cooled relative to our reference atmosphere. There is no water opacity driven heating as was found in the tidally-locked models of \citetalias{2024arXiv240911151S}. 
	\item There are two possible causes of this cooling: i) radiative cooling by water vapour between the tropopause and stratopause, and (ii) a decrease in ozone abundance in the stratosphere and mesosphere. 
	\item The photolysis of impact-delivered water leads to up to an order of magnitude increase in the abundance of hydroxyl (\ce{OH}) and hydroperoxyl (\ce{HO2}) radicals post-impact, both of which form part of a catalytic cycle destroying atmospheric ozone. 
	\item We find up to an order of magnitude depletion in ozone for pressures between $10^{-7}<P<10^{-3}$ bar, with the relative destruction peaking at $\sim10^{-5}$ bar (i.e., the mesopause). However, note that since this destruction mostly occurs at low pressures, the mean change in the total ozone column density over the first three years post-impact is small, less than two Dobson Units (one Dobson Unit is $2.69\times10^{20}\,\mathrm{mol\,m^{-2}}$), or $<1\%$ Earth's ozone column density.
	\item We find small, impact-induced enrichments in the abundances of hydrogen- and oxygen-rich molecules, such as \ce{H2}, \ce{O2}, \ce{OH}, \ce{HO2}, \ce{NO} and \ce{NO2} which persist to quasi-steady-state.
	\item The impact-induced changes to both the transmission and emission spectra are small, typically falling far below the detection threshold of JWST. In particular, we do not find evidence of the post-impact increase in apparent radius seen by \citetalias{2024arXiv240911151S}.
	\item This again can be linked to the weaker vertical advection in our exo-Earth-atmosphere driving a weaker enhancement of water vapour, and hence (light scattering) cloud ice formation at low pressures. 
	\item However cloud ice formation may have an effect on the reflectivity of the planet (i.e., the albedo). This change, paired with the water-vapour driven changes in absorption feature strength, represents the best opportunity for detecting an individual massive impact with an exo-Earth's atmosphere. But even then, the effects of seasonality and long timescale atmospheric trends may  mask these changes.
\end{itemize}

Comparing our results with the tidally-locked model of \citetalias{2024arXiv240911151S}, it is clear the global atmospheric circulations play a key role in setting the post-impact observability of a single, massive, cometary impact. Only for a tidally-locked atmosphere with a global overturning circulation do we find that enough water is carried aloft to drive a potentially observable, cloud ice driven, scattering of light during transit. Combined with the expected rarity of impacts for terrestrial worlds (see, for example, \citealt{10.1046/j.1365-8711.2000.03568.x}), it is thus unlikely that any individual impacts will be observed. \\
Instead the most important { effects} associated with an individual cometary impact are likely to be the long-lasting changes to atmospheric composition and chemistry which occur. These changes suggest that impacts with younger planets, where we also expect the impact rate to be significantly higher (as it was for the Earth; e.g. \citealt{Fassett2013,2020AsBio..20.1121O}), may have played an important role in delivering volatiles (e.g. \citealt{OWEN1995215,2002ESASP.518....9E,2018SoSyR..52..392M,2023PhyU...66....2M}) and hence oxygenating their initially oxygen-poor atmospheres (like that of the Archean Earth; see the review by, for example, \citealt{annurev:/content/journals/10.1146/annurev.earth.33.092203.122711} or \citealt{Kump2008}). We will explore this in our next major study, using our coupled impact and climate model to study the effects of repeated cometary bombardment in, initially, oxygen-poor exo-Earth analogue atmospheres. 

\begin{acknowledgements}
\nolinenumbers
F. Sainsbury-Martinez and C. Walsh would like to thank UK Research and Innovation for support under grant number MR/T040726/1. Additionally, C. Walsh would like to thank the University of Leeds and the Science and Technology Facilities Council for financial support (ST/X001016/1). This work was undertaken on ARC4, part of the High Performance Computing facilities at the University of Leeds, UK.\\
\end{acknowledgements}

\appendix

\begin{figure*}[tbp]
\begin{centering}
\includegraphics[width=0.85\textwidth]{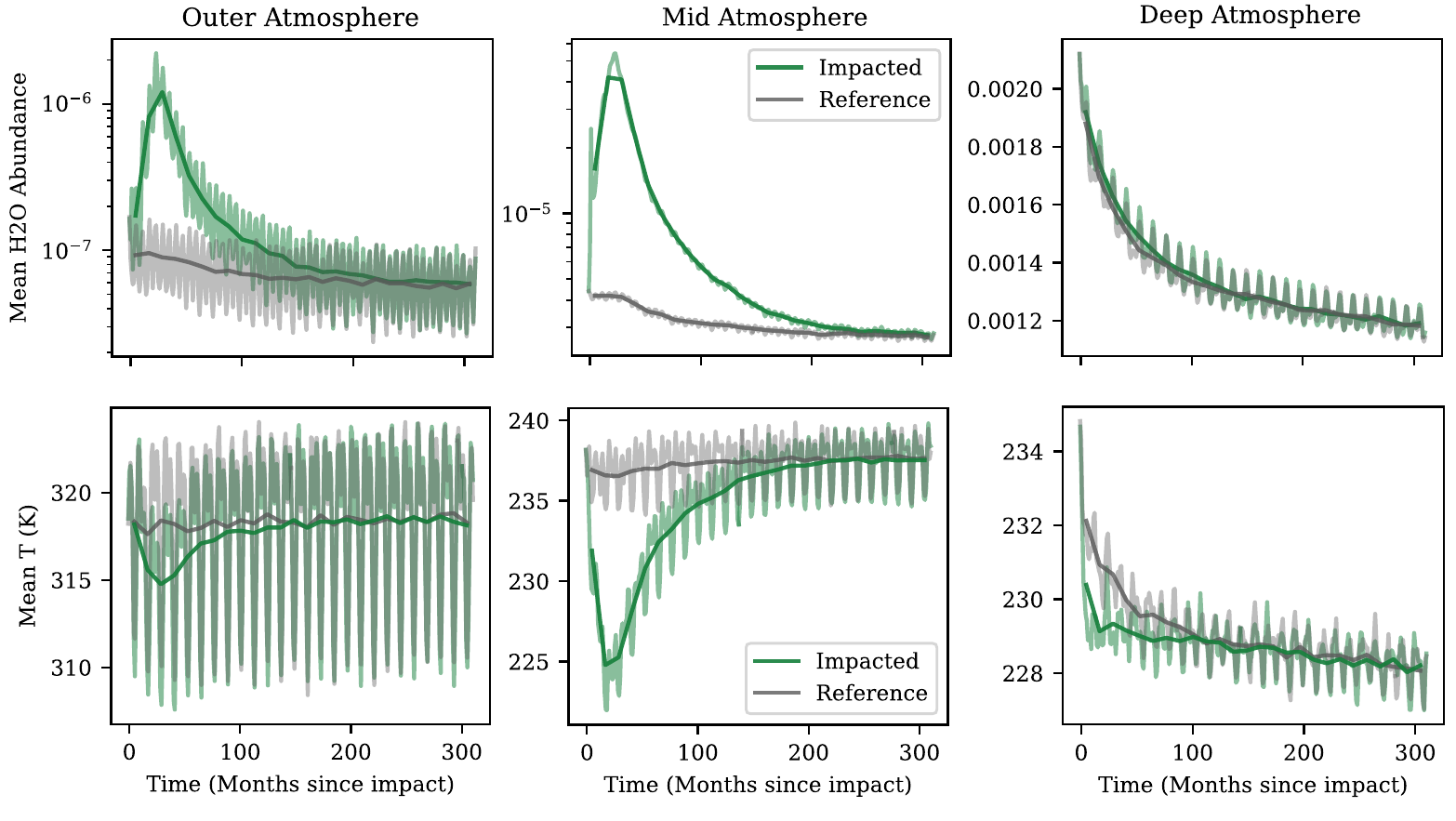}
\caption{ Time evolution of the annual mean (opaque lines) and monthly mean (faint lines) fractional water abundance (top row) and temperature (bottom row) in the outer atmosphere ($P<10^{-5}$ bar - left), mid-atmosphere ($10^{-5}>P>10^{-2}$ bar - middle), and near the surface ($P>10^{-2}$ bar - right) for both our exo-Earth analogue cometary impact model (green) and reference atmosphere (grey). Here we plot data for $25$ years, extending beyond the $\sim20$ years it takes for our model to reach a quasi-steady-state for validation purposes.   \label{fig:combined_time_evolution} }
\end{centering}
\end{figure*}

\begin{figure}[tbp]
\begin{centering}
\includegraphics[width=0.5\textwidth]{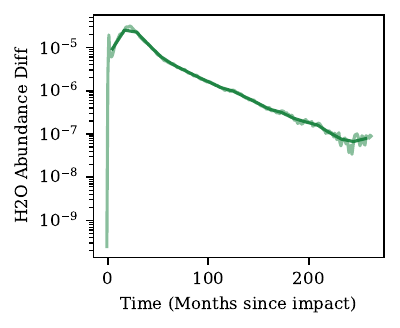}
\caption{{ Time evolution of the annual mean (opaque lines) and monthly mean (faint lines) difference in fractional water abundance between our impacted and reference exo-Earth analogue atmospheres. Here we consider the mean difference in water abundance in the mid-atmosphere ($10^{-5}>P>10^{-2}$ bar) as this is the region in which the impact-induced changes to the fractional water abundance last longest. Note that we consider the model to have reached a quasi-steady-state when the difference reaches a minima, which occurs around 20 years post-impact.  \label{fig:MA_water_diff}} }
\end{centering}
\end{figure}

\begin{figure*}[tbp]
\begin{centering}
\includegraphics[width=0.85\textwidth]{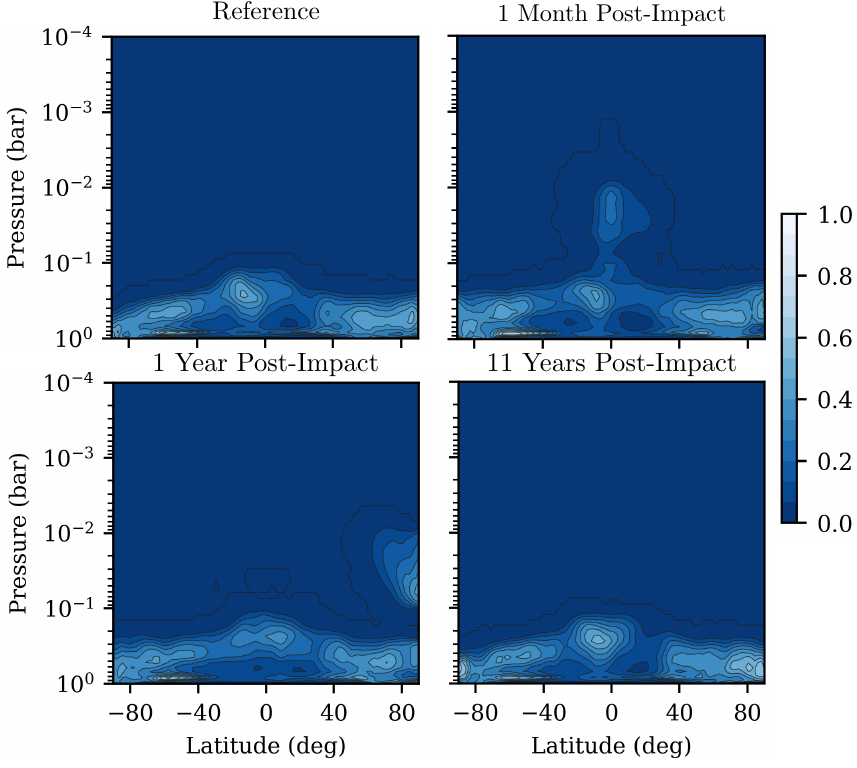}
\caption{ Zonally and temporally averaged cloud fraction for both our reference atmosphere (top left) and at three points in time in our exo-Earth analogue cometary impact model: one month (top right), one year (bottom left), and eleven years (bottom right) post-impact, by which time the cloud fraction has returned to a state similar to that found in our non-impacted reference atmosphere. Note that here we have chosen to limit our plots to pressures $P>10^{-4}$ bar due to the limited pressure range over which CESM's cloud model is enabled ($P>10^{-3}$ bar).   \label{fig:cloud_fraction} }
\end{centering}
\end{figure*}

\begin{figure*}[tbp]
\begin{centering}
\includegraphics[width=0.85\textwidth]{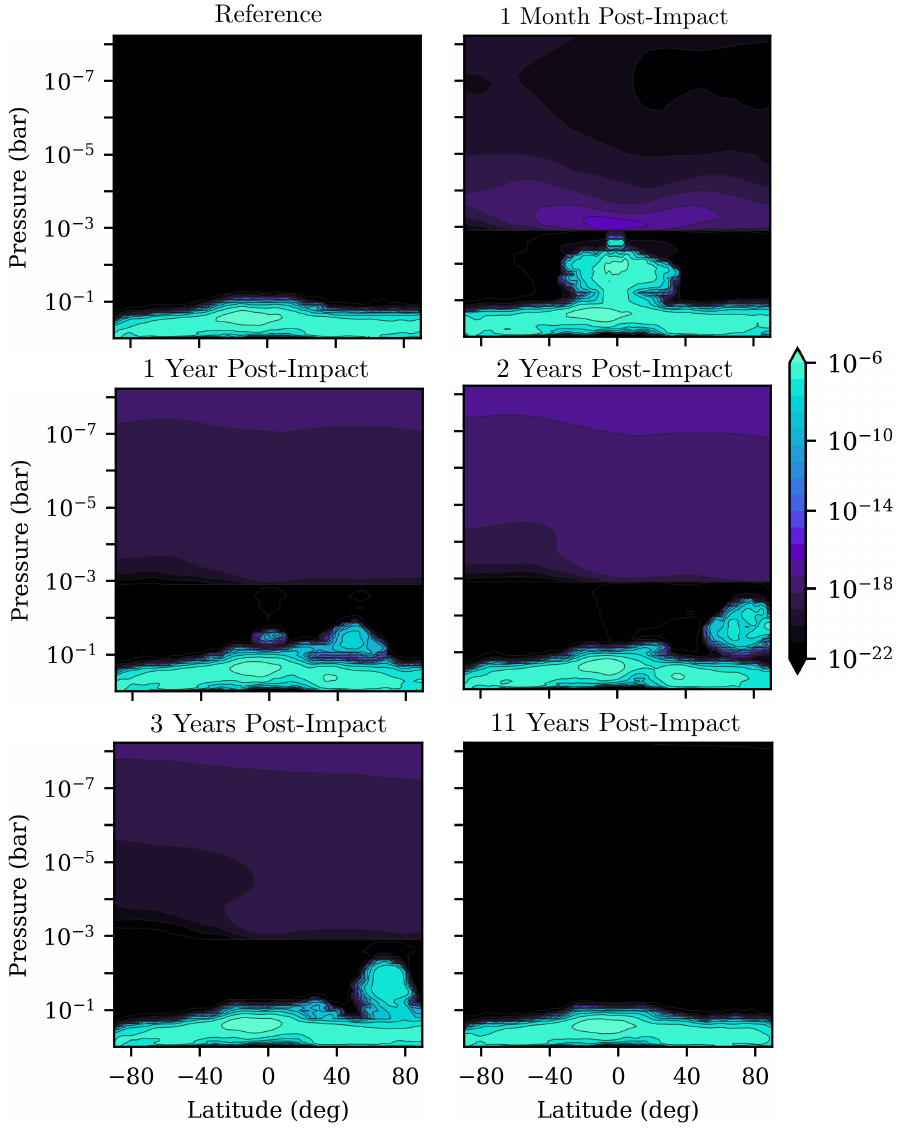}
\caption{Zonally and temporally averaged cloud ice mixing ratio for both our reference atmosphere (top left) and at five points in time in our exo-Earth analogue cometary impact model: one month (top right), one year (middle left), two years (middle right), three years (bottom left), and eleven years (bottom right) post-impact, by which time the average cloud ice mixing ratio has returned to a state similar to that found in our reference atmosphere.   \label{fig:cloud_ice} }
\end{centering}
\end{figure*}
\section{Appendix} \label{sec:appendix}
To aid in the comparison of our exo-Earth analogue model with the tidally-locked model of \citetalias{2024arXiv240911151S} we reproduce three of their figures using data from our exo-Earth analogue model: i) \autoref{fig:combined_time_evolution} shows the time-evolution of the mean fractional water abundance and temperature in the outer ($P<10^{-5}$ bar), mid ($10^{-2}<P<10^{-5}$ bar), and deep (near-surface; $P>10^{-2}$ bar) atmosphere (to compare with their figure 3); ii) \autoref{fig:cloud_fraction} shows the temporally and zonally averaged cloud fraction at three points in time post-impact, comparing them with our reference atmosphere (to compare with their figure 10); and iii) \autoref{fig:cloud_ice} shows the zonally and temporally averaged cloud ice mixing ratio at five points in time post impact, again comparing with our reference atmosphere (to compare with their figure 12). \\

The time evolution profiles which we plot in \autoref{fig:combined_time_evolution} reveal both the delayed advection of water to the outer atmosphere as well as the cooling effect of the impact-delivered material for all pressure levels, but particularly for the mid-atmosphere. 
Here, we find a peak difference of $17\,$K between the annual mean temperature of the impacted and reference atmospheres. We also find that this cooling, as well as the cooling of the outer and deep atmosphere, is longer lasting than the changes found by \citetalias{2024arXiv240911151S}. 
The most likely culprit is a difference in what is driving the temperature change. In \citetalias{2024arXiv240911151S}, the heating is driven by the local opacity of water deposited at the sub-stellar point, water which is rapidly mixed throughout the atmosphere and away from the heat source. On the other hand, the changes seen here are associated with water-induced changes to the global climate, including radiative cooling by water vapour and destruction of low pressure ozone leading to a reduction in stratospheric and mesospheric heating. Note that similar effects may also have been at play in the tidally-locked atmosphere of \citetalias{2024arXiv240911151S} but masked by the strong sub-stellar point heating. \\
{ We also plot, in \autoref{fig:MA_water_diff}, the evolution of the difference in mean mid-atmosphere water abundance between our impacted and reference exo-Earth analogue atmospheres. This difference reaches a minima approximately twenty years post-impact, at which point we consider the atmosphere to have reached a quasi-steady-state. \\ }

Moving on to the mean cloud fraction (\autoref{fig:cloud_fraction}), we find that the impact affects the cloud profile in two ways. Shortly after the impact (top right), we find that a column of cloud forms at the equator. However irradiation by the host star means that this high altitude equatorial cloud does not persist. Instead, one year post-impact (bottom left), we find that the deposited water vapour leads to cloud formation over the north pole between $10^{-2}$ and $10^{-1}$ bar. Comparing this cloud fraction profile with \citetalias{2024arXiv240911151S}, we find two major differences: weaker cloud formation over the pole and an asymmetry between the cloud coverage over the poles. Both of these effects can be linked with the seasonality of our exo-Earth analogue. Clouds form from atmospheric water vapour over the poles in winter, hence the cloud formation in our February snapshot of the atmosphere. Then, in the summer, when the illumination of the poles increases, they burn-off, leading to the asymmetric profile seen here. This also both prevents the formation of the very high cloud fractions found by \citetalias{2024arXiv240911151S}, whose tidally-locked atmosphere has permanently cold and dark poles in which clouds can accumulate, and explains why the cloud coverage profile rapidly returns towards a state reminiscent of our reference atmosphere. \\

Finally we come to the mean cloud ice mixing ratio (\autoref{fig:cloud_ice}), which we plot at five points in time so as to better demonstrate the role that vertical advection plays in cloud ice formation. 
Briefly, for our exo-Earth analogue atmosphere, we find that one month post-impact (top right), most of the cloud ice formation is confined to the same pressure levels at which the material from the cometary breakup is deposited (i.e. $P>5\times10^{-3}$ bar). However there is also a slight hint of ice formation at lower pressures, particularly just above the impact site, which we attribute to freeze out of water deposited during the thermal ablation of the comet as well some weak vertical advection from the breakup site. With time, this vertical advection of water does lead to the formation of low-pressure cloud ice. However, comparing our results with \citetalias{2024arXiv240911151S}, we find that the abundance of this low-pressure cloud ice is so much weaker that we have even had to reduce the scale of our plots to maintain a useful dynamic range. For example, if we compare profiles taken when the water content, and hence, the cloud ice content, of the outer atmosphere is at its highest (two years post-impact here versus approximately two months post-impact in \citetalias{2024arXiv240911151S}), we find that the cloud ice mixing ratio in our exo-Earth analogue remains almost negligible. This can explain the difference in transmission spectra between the two models. In our post-impact exo-Earth analogue atmosphere, there is not enough cloud ice formed to scatter light passing through the outer atmosphere and hence, increase the apparent radius of the planet.

\bibliography{papers}{}

\begin{thebibliography}{}
\expandafter\ifx\csname natexlab\endcsname\relax\def\natexlab#1{#1}\fi
\providecommand{\url}[1]{\href{#1}{#1}}
\providecommand{\dodoi}[1]{doi:~\href{http://doi.org/#1}{\nolinkurl{#1}}}
\providecommand{\doeprint}[1]{\href{http://ascl.net/#1}{\nolinkurl{http://ascl.net/#1}}}
\providecommand{\doarXiv}[1]{\href{https://arxiv.org/abs/#1}{\nolinkurl{https://arxiv.org/abs/#1}}}

\bibitem[{{Alibert} {et~al.}(2005){Alibert}, {Mordasini}, {Benz}, \&
  {Winisdoerffer}}]{2005A&A...434..343A}
{Alibert}, Y., {Mordasini}, C., {Benz}, W., \& {Winisdoerffer}, C. 2005, \aap,
  434, 343, \dodoi{10.1051/0004-6361:20042032}

\bibitem[{Anders(1989)}]{anders1989}
Anders, E. 1989, Nature, 342, 255

\bibitem[{{Anslow} {et~al.}(2023){Anslow}, {Bonsor}, \&
  {Rimmer}}]{2023RSPSA.47930434A}
{Anslow}, R.~J., {Bonsor}, A., \& {Rimmer}, P.~B. 2023, Proceedings of the
  Royal Society of London Series A, 479, 20230434,
  \dodoi{10.1098/rspa.2023.0434}

\bibitem[{{Anslow} {et~al.}(2025){Anslow}, {Bonsor}, {Todd}, {Wordsworth},
  {Rae}, {McDonald}, \& {Rimmer}}]{2025MNRAS.539..376A}
{Anslow}, R.~J., {Bonsor}, A., {Todd}, Z.~R., {et~al.} 2025, \mnras, 539, 376,
  \dodoi{10.1093/mnras/staf507}

\bibitem[{{Bhongade} {et~al.}(2024){Bhongade}, {Marsh}, {Sainsbury-Martinez},
  \& {Cooke}}]{anand2024}
{Bhongade}, A., {Marsh}, D.~R., {Sainsbury-Martinez}, F., \& {Cooke}, G.~J.
  2024, arXiv e-prints, arXiv:2407.02444, \dodoi{10.48550/arXiv.2407.02444}

\bibitem[{{Bockel{\'e}e-Morvan} {et~al.}(2004){Bockel{\'e}e-Morvan},
  {Crovisier}, {Mumma}, \& {Weaver}}]{2004come.book..391B}
{Bockel{\'e}e-Morvan}, D., {Crovisier}, J., {Mumma}, M.~J., \& {Weaver}, H.~A.
  2004, in Comets II, ed. M.~C. {Festou}, H.~U. {Keller}, \& H.~A. {Weaver},
  391

\bibitem[{Canfield(2005)}]{annurev:/content/journals/10.1146/annurev.earth.33.092203.122711}
Canfield, D. 2005, Annual Review of Earth and Planetary Sciences, 33, 1,
  \dodoi{https://doi.org/10.1146/annurev.earth.33.092203.122711}

\bibitem[{Chyba \& Sagan(1992)}]{chyba1992}
Chyba, C., \& Sagan, C. 1992, Nature, 355, 125

\bibitem[{Chyba {et~al.}(1990)Chyba, Thomas, Brookshaw, \& Sagan}]{11538074}
Chyba, C.~F., Thomas, P.~J., Brookshaw, L., \& Sagan, C. 1990, Science, 249,
  366, \dodoi{10.1126/science.11538074}

\bibitem[{Cooke {et~al.}(2022)Cooke, Marsh, Walsh, Rugheimer, \&
  Villanueva}]{10.1093/mnras/stac2604}
Cooke, G.~J., Marsh, D.~R., Walsh, C., Rugheimer, S., \& Villanueva, G.~L.
  2022, Monthly Notices of the Royal Astronomical Society, 518, 206,
  \dodoi{10.1093/mnras/stac2604}

\bibitem[{{Cooke} {et~al.}(2024){Cooke}, {Marsh}, {Walsh}, \&
  {Sainsbury-Martinez}}]{cooke2024}
{Cooke}, G.~J., {Marsh}, D.~R., {Walsh}, C., \& {Sainsbury-Martinez}, F. 2024,
  \psj, 5, 168, \dodoi{10.3847/PSJ/ad53c3}

\bibitem[{{Cooke} {et~al.}(2023){Cooke}, {Marsh}, {Walsh}, \&
  {Youngblood}}]{2023ApJ...959...45C}
{Cooke}, G.~J., {Marsh}, D.~R., {Walsh}, C., \& {Youngblood}, A. 2023, \apj,
  959, 45, \dodoi{10.3847/1538-4357/ad0381}

\bibitem[{{Davoudi} {et~al.}(2024){Davoudi}, {Rackham}, {Gillon}, {de Wit},
  {Burgasser}, {Delrez}, {Iyer}, \& {Ducrot}}]{2024ApJ...970L...4D}
{Davoudi}, F., {Rackham}, B.~V., {Gillon}, M., {et~al.} 2024, \apjl, 970, L4,
  \dodoi{10.3847/2041-8213/ad5c6c}

\bibitem[{{de F. Forster} \& {Shine}(1999)}]{1999GeoRL..26.3309D}
{de F. Forster}, P.~M., \& {Shine}, K.~P. 1999, \grl, 26, 3309,
  \dodoi{10.1029/1999GL010487}

\bibitem[{Delsemme(2000)}]{DELSEMME2000313}
Delsemme, A.~H. 2000, Icarus, 146, 313,
  \dodoi{https://doi.org/10.1006/icar.2000.6404}

\bibitem[{{Ehrenfreund} {et~al.}(2002){Ehrenfreund}, {Irvine}, {Becker},
  {Blank}, {Brucato}, {Colangeli}, {Derenne}, {Despois}, {Dutrey}, {Fraaije},
  {Lazcano}, {Owen}, {Robert}, \& {International Space Science Institute
  Issi-Team}}]{2002ESASP.518....9E}
{Ehrenfreund}, P., {Irvine}, W., {Becker}, L., {et~al.} 2002, in ESA Special
  Publication, Vol. 518, Exo-Astrobiology, ed. H.~{Lacoste}, 9--14

\bibitem[{Fassett \& Minton(2013)}]{Fassett2013}
Fassett, C.~I., \& Minton, D.~A. 2013, Nature Geoscience, 6, 520,
  \dodoi{10.1038/ngeo1841}

\bibitem[{{Frantseva} {et~al.}(2020){Frantseva}, {Mueller}, {Pokorn{\'y}}, {van
  der Tak}, \& {ten Kate}}]{2020A&A...638A..50F}
{Frantseva}, K., {Mueller}, M., {Pokorn{\'y}}, P., {van der Tak}, F.~F.~S., \&
  {ten Kate}, I.~L. 2020, \aap, 638, A50, \dodoi{10.1051/0004-6361/201936783}

\bibitem[{Gettelman {et~al.}(2019)Gettelman, Mills, Kinnison, Garcia, Smith,
  Marsh, Tilmes, Vitt, Bardeen, McInerny, Liu, Solomon, Polvani, Emmons,
  Lamarque, Richter, Glanville, Bacmeister, Phillips, Neale, Simpson, DuVivier,
  Hodzic, \& Randel}]{https://doi.org/10.1029/2019JD030943}
Gettelman, A., Mills, M.~J., Kinnison, D.~E., {et~al.} 2019, Journal of
  Geophysical Research: Atmospheres, 124, 12380,
  \dodoi{https://doi.org/10.1029/2019JD030943}

\bibitem[{{Grenfell} {et~al.}(2006){Grenfell}, {Stracke}, {Patzer}, {Titz}, \&
  {Rauer}}]{2006IJAsB...5..295G}
{Grenfell}, J.~L., {Stracke}, B., {Patzer}, B., {Titz}, R., \& {Rauer}, H.
  2006, International Journal of Astrobiology, 5, 295,
  \dodoi{10.1017/S1473550406003478}

\bibitem[{{Hammond} \& {Lewis}(2021)}]{2021PNAS..11822705H}
{Hammond}, M., \& {Lewis}, N.~T. 2021, Proceedings of the National Academy of
  Science, 118, e2022705118, \dodoi{10.1073/pnas.2022705118}

\bibitem[{Hughes(2000)}]{10.1046/j.1365-8711.2000.03568.x}
Hughes, D.~W. 2000, Monthly Notices of the Royal Astronomical Society, 317,
  429, \dodoi{10.1046/j.1365-8711.2000.03568.x}

\bibitem[{{Ipatov} \& {Mather}(2003)}]{2003EM&P...92...89I}
{Ipatov}, S.~I., \& {Mather}, J.~C. 2003, Earth Moon and Planets, 92, 89,
  \dodoi{10.1023/B:MOON.0000031928.45965.7b}

\bibitem[{{Ipatov} \& {Mather}(2004)}]{2004AdSpR..33.1524I}
---. 2004, Advances in Space Research, 33, 1524,
  \dodoi{10.1016/S0273-1177(03)00451-4}

\bibitem[{{Komacek} \& {Abbot}(2019)}]{2019ApJ...871..245K}
{Komacek}, T.~D., \& {Abbot}, D.~S. 2019, \apj, 871, 245,
  \dodoi{10.3847/1538-4357/aafb33}

\bibitem[{Kump(2008)}]{Kump2008}
Kump, L.~R. 2008, Nature, 451, 277, \dodoi{10.1038/nature06587}

\bibitem[{{Le Roy} {et~al.}(2015){Le Roy}, {Altwegg}, {Balsiger}, {Berthelier},
  {Bieler}, {Briois}, {Calmonte}, {Combi}, {De Keyser}, {Dhooghe}, {Fiethe},
  {Fuselier}, {Gasc}, {Gombosi}, {H{\"a}ssig}, {J{\"a}ckel}, {Rubin}, \&
  {Tzou}}]{2015AA...583A...1L}
{Le Roy}, L., {Altwegg}, K., {Balsiger}, H., {et~al.} 2015, \aap, 583, A1,
  \dodoi{10.1051/0004-6361/201526450}

\bibitem[{{Lin} \& {Kaltenegger}(2022)}]{2022MNRAS.516.3167L}
{Lin}, Z., \& {Kaltenegger}, L. 2022, \mnras, 516, 3167,
  \dodoi{10.1093/mnras/stac2536}

\bibitem[{{Liu} {et~al.}(2023){Liu}, {Marsh}, {Walsh}, \&
  {Cooke}}]{2023MNRAS.524.1491L}
{Liu}, B., {Marsh}, D.~R., {Walsh}, C., \& {Cooke}, G. 2023, \mnras, 524, 1491,
  \dodoi{10.1093/mnras/stad1828}

\bibitem[{{Marov} \& {Ipatov}(2018)}]{2018SoSyR..52..392M}
{Marov}, M.~Y., \& {Ipatov}, S.~I. 2018, Solar System Research, 52, 392,
  \dodoi{10.1134/S0038094618050052}

\bibitem[{{Marov} \& {Ipatov}(2023)}]{2023PhyU...66....2M}
---. 2023, Physics Uspekhi, 66, 2, \dodoi{10.3367/UFNe.2021.08.039044}

\bibitem[{{McClure} {et~al.}(2023){McClure}, {Rocha}, {Pontoppidan}, {Crouzet},
  {Chu}, {Dartois}, {Lamberts}, {Noble}, {Pendleton}, {Perotti}, {Qasim},
  {Rachid}, {Smith}, {Sun}, {Beck}, {Boogert}, {Brown}, {Caselli}, {Charnley},
  {Cuppen}, {Dickinson}, {Drozdovskaya}, {Egami}, {Erkal}, {Fraser}, {Garrod},
  {Harsono}, {Ioppolo}, {Jim{\'e}nez-Serra}, {Jin}, {J{\o}rgensen},
  {Kristensen}, {Lis}, {McCoustra}, {McGuire}, {Melnick}, {{\~A}-berg},
  {Palumbo}, {Shimonishi}, {Sturm}, {van Dishoeck}, \&
  {Linnartz}}]{2023NatAs...7..431M}
{McClure}, M.~K., {Rocha}, W.~R.~M., {Pontoppidan}, K.~M., {et~al.} 2023,
  Nature Astronomy, 7, 431, \dodoi{10.1038/s41550-022-01875-w}

\bibitem[{{Mordasini} {et~al.}(2016){Mordasini}, {van Boekel}, {Molli{\`e}re},
  {Henning}, \& {Benneke}}]{2016ApJ...832...41M}
{Mordasini}, C., {van Boekel}, R., {Molli{\`e}re}, P., {Henning}, T., \&
  {Benneke}, B. 2016, \apj, 832, 41, \dodoi{10.3847/0004-637X/832/1/41}

\bibitem[{{O'Brien} {et~al.}(2014){O'Brien}, {Walsh}, {Morbidelli}, {Raymond},
  \& {Mandell}}]{2014Icar..239...74O}
{O'Brien}, D.~P., {Walsh}, K.~J., {Morbidelli}, A., {Raymond}, S.~N., \&
  {Mandell}, A.~M. 2014, \icarus, 239, 74, \dodoi{10.1016/j.icarus.2014.05.009}

\bibitem[{Or{\'o}(1961)}]{oro1961}
Or{\'o}, J. 1961, Nature, 190, 389, \dodoi{10.1038/190389a0}

\bibitem[{{Osinski} {et~al.}(2020){Osinski}, {Cockell}, {Pontefract}, \&
  {Sapers}}]{2020AsBio..20.1121O}
{Osinski}, G.~R., {Cockell}, C.~S., {Pontefract}, A., \& {Sapers}, H.~M. 2020,
  Astrobiology, 20, 1121, \dodoi{10.1089/ast.2019.2203}

\bibitem[{Owen \& Bar-Nun(1995)}]{OWEN1995215}
Owen, T., \& Bar-Nun, A. 1995, Icarus, 116, 215,
  \dodoi{https://doi.org/10.1006/icar.1995.1122}

\bibitem[{Passey \& Melosh(1980)}]{PASSEY1980211}
Passey, Q.~R., \& Melosh, H. 1980, Icarus, 42, 211,
  \dodoi{https://doi.org/10.1016/0019-1035(80)90072-X}

\bibitem[{{Pierrehumbert} \& {Hammond}(2019)}]{2019AnRFM..51..275P}
{Pierrehumbert}, R.~T., \& {Hammond}, M. 2019, Annual Review of Fluid
  Mechanics, 51, 275, \dodoi{10.1146/annurev-fluid-010518-040516}

\bibitem[{Ribas {et~al.}(2005)Ribas, Guinan, Güdel, \& Audard}]{Ribas_2005}
Ribas, I., Guinan, E.~F., Güdel, M., \& Audard, M. 2005, The Astrophysical
  Journal, 622, 680, \dodoi{10.1086/427977}

\bibitem[{{Rugheimer} \& {Kaltenegger}(2018)}]{2018ApJ...854...19R}
{Rugheimer}, S., \& {Kaltenegger}, L. 2018, \apj, 854, 19,
  \dodoi{10.3847/1538-4357/aaa47a}

\bibitem[{{Sainsbury-Martinez} \& {Walsh}(2024)}]{2024ApJ...966...39S}
{Sainsbury-Martinez}, F., \& {Walsh}, C. 2024, \apj, 966, 39,
  \dodoi{10.3847/1538-4357/ad28b3}

\bibitem[{{Sainsbury-Martinez} {et~al.}(2025){Sainsbury-Martinez}, {Walsh}, \&
  {Cooke}}]{2024arXiv240911151S}
{Sainsbury-Martinez}, F., {Walsh}, C., \& {Cooke}, G. 2025, \apj, 982, 29,
  \dodoi{10.3847/1538-4357/ad96ad}

\bibitem[{{Sainsbury-Martinez} {et~al.}(2024){Sainsbury-Martinez}, {Walsh},
  {Cooke}, \& {Marsh}}]{sainsbury2024b}
{Sainsbury-Martinez}, F., {Walsh}, C., {Cooke}, G.~J., \& {Marsh}, D.~R. 2024,
  arXiv e-prints, arXiv:2407.01480, \dodoi{10.48550/arXiv.2407.01480}

\bibitem[{{Sasselov} {et~al.}(2020){Sasselov}, {Grotzinger}, \&
  {Sutherland}}]{2020SciA....6.3419S}
{Sasselov}, D.~D., {Grotzinger}, J.~P., \& {Sutherland}, J.~D. 2020, Science
  Advances, 6, eaax3419, \dodoi{10.1126/sciadv.aax3419}

\bibitem[{{Selsis}(2000)}]{2000ESASP.451..133S}
{Selsis}, F. 2000, in ESA Special Publication, Vol. 451, Darwin and Astronomy :
  the Infrared Space Interferometer, ed. B.~{Sch{\"u}rmann}, 133

\bibitem[{Snodgrass {et~al.}(2017)Snodgrass, Agarwal, Combi, Fitzsimmons,
  Guilbert-Lepoutre, Hsieh, Hui, Jehin, Kelley, Knight, Opitom, Orosei,
  de~Val-Borro, \& Yang}]{snodgrass2017}
Snodgrass, C., Agarwal, J., Combi, M., {et~al.} 2017, The Astronomy and
  Astrophysics Review, 25, 5

\bibitem[{Stocker {et~al.}(2024)Stocker, Steiner, Ladst{\"a}dter, Foelsche, \&
  Randel}]{Stocker2024}
Stocker, M., Steiner, A.~K., Ladst{\"a}dter, F., Foelsche, U., \& Randel, W.~J.
  2024, Communications Earth {\&} Environment, 5, 450,
  \dodoi{10.1038/s43247-024-01620-3}

\bibitem[{{Svetsov} {et~al.}(1995){Svetsov}, {Nemtchinov}, \&
  {Teterev}}]{1995Icar..116..131S}
{Svetsov}, V.~V., {Nemtchinov}, I.~V., \& {Teterev}, A.~V. 1995, \icarus, 116,
  131, \dodoi{10.1006/icar.1995.1116}

\bibitem[{Todd \& \"{O}berg(2020)}]{doi:10.1089/ast.2019.2187}
Todd, Z.~R., \& \"{O}berg, K.~I. 2020, Astrobiology, 20, 1109,
  \dodoi{10.1089/ast.2019.2187}

\bibitem[{{Villanueva} {et~al.}(2018){Villanueva}, {Smith}, {Protopapa},
  {Faggi}, \& {Mandell}}]{2018JQSRT.217...86V}
{Villanueva}, G.~L., {Smith}, M.~D., {Protopapa}, S., {Faggi}, S., \&
  {Mandell}, A.~M. 2018, \jqsrt, 217, 86, \dodoi{10.1016/j.jqsrt.2018.05.023}

\bibitem[{Waajen {et~al.}(2024)Waajen, Lima, Goodacre, \& Cockell}]{Waajen2024}
Waajen, A.~C., Lima, C., Goodacre, R., \& Cockell, C.~S. 2024, Scientific
  Reports, 14, 3691

\bibitem[{Wennberg {et~al.}(1994)Wennberg, Cohen, Stimpfle, Koplow, Anderson,
  Salawitch, Fahey, Woodbridge, Keim, Gao, Webster, May, Toohey, Avallone,
  Proffitt, Loewenstein, Podolske, Chan, \&
  Wofsy}]{doi:10.1126/science.266.5184.398}
Wennberg, P.~O., Cohen, R.~C., Stimpfle, R.~M., {et~al.} 1994, Science, 266,
  398, \dodoi{10.1126/science.266.5184.398}

\end{thebibliography}
\bibliographystyle{aasjournal}

\end{document}